\documentclass [12pt,eqsecnum,amsfonts,amssymb,aps]{revtex4}

\input epsf

\usepackage[usenames]{color}
\usepackage{graphicx}
\usepackage{bm}
\usepackage{rotating}

\newcommand{\beq}{\begin{equation}}
\newcommand{\eeq}{\end{equation}}
\newcommand{\beqs}{\begin{eqnarray}}
\newcommand{\eeqs}{\end{eqnarray}}

\begin{document}

\baselineskip 6.0mm

\title{Asymptotic Behavior of Acyclic and Cyclic Orientations of 
Directed Lattice Graphs} 

\author{Shu-Chiuan Chang$^a$ and Robert Shrock$^b$}

\affiliation{(a) \ Department of Physics, National Cheng Kung University,
Tainan 70101, Taiwan} 

\affiliation{(b) \ C. N. Yang Institute for Theoretical Physics and
Department of Physics and Astronomy \\
Stony Brook University, Stony Brook, NY 11794, USA }

\begin{abstract}

  We calculate exponential growth constants describing the asymptotic behavior
  of several quantities enumerating classes of orientations of arrow variables
  on the bonds of several types of directed lattice strip graphs $G$ of finite
  width and arbitrarily great length, in the infinite-length limit, denoted
  $\{G\}$.  Specifically, we calculate the exponential growth constants for (i)
  acyclic orientations, $\alpha(\{G\})$, (ii) acyclic orientations with a
  single source vertex, $\alpha_0(\{G\})$, and (iii) totally cyclic
  orientations, $\beta(\{G\})$. We consider several lattices, including square
  ($sq$), triangular ($tri$), and honeycomb ($hc$). From our calculations, we
  infer lower and upper bounds on these exponential growth constants for the
  respective infinite lattices. To our knowledge, these are the best current
  bounds on these quantities. Since our lower and upper bounds are quite close
  to each other, we can infer very accurate approximate values for the
  exponential growth constants, with fractional uncertainties ranging from
  $O(10^{-4})$ to $O(10^{-2})$.  Further, we present exact values of
  $\alpha(tri)$, $\alpha_0(tri)$, and $\beta(hc)$ and use them to show that our
  lower and upper bounds on these quantities are very close to these exact
  values, even for modest strip widths. Results are also given for a nonplanar
  lattice denoted $sq_d$. We show that $\alpha(\{G\})$, $\alpha_0(\{ G \})$,
  and $\beta(\{G\})$ are monotonically increasing functions of vertex degree
  for these lattices. We also study the asymptotic behavior of the ratios of
  the quantities (i)-(iii) divided by the total number of edge orientations as
  the number of vertices goes to infinity.  A comparison is given of these
  exponential growth constants with the corresponding exponential growth
  constant $\tau(\{ G\})$ for spanning trees. Our results are in agreement with
  inequalities following from the Merino-Welsh and Conde-Merino conjectures.

\end{abstract}

\maketitle


\pagestyle{plain}
\pagenumbering{arabic}


\section{Introduction and Basics}
\label{intro_section}

In this paper we report new results on three quantities defined on directed
graphs, namely acyclic orientations, acyclic orientations with a single source
vertex, and totally cyclic orientations of families of directed graphs.  We
present calculations of the exponential growth constants for these quantities
for strip graphs of several lattices in the limit of infinite strip length.
From these calculations we infer lower and upper bounds on the three types of
exponential growth constants on the thermodynamic limits of the respective
lattices. These are quite close to each other, even for modest maximal values
of strip width used.  Hence, we are able to infer the values of these growth
constants to high accuracies, with fractional uncertainties ranging from the
$O(10^{-4})$ to $O(10^{-2})$.  We also present exact results for three
exponential growth constants.  From our calculations, we show that the
exponential growth constants for acyclic orientations, acyclic orientations
with a single source, and totally cyclic orientations are monotonically
increasing functions of vertex degree. Comparisons are also made with the
exponential growth constants for spanning trees on these lattices. The results
are in agreement with inequalities implied by the Merino-Welsh and the
Conde-Merino conjectures.

We begin with some basic background and definitions. Let $G=(V,E)$ be a graph
defined by its vertex and edge sets $V$ and $E$.  Let $n(G)=|V|$, $e(G)=|E|$,
and $k(G)$ denote the number of vertices, edges (= bonds), and connected
components of $G$, respectively.  We will often use the simpler symbol $n
\equiv n(G)$ where no confusion will result. Without loss of generality, we
restrict our analysis to connected graphs here, so $k(G)=1$.  We also 
restrict our analysis to graphs that do not contain loops (i.e., edges that
connect a vertex to itself) or multiple edges joining a given pair of
vertices. The reasons for this exclusion will be explained below. The
degree $\Delta(v_i)$ of a vertex $v_i \in V$ is the number of edges that are
incident on $v_i$.  Given a graph $G$, one can assign an arrow to each edge of
$G$, thereby defining a directed graph, also called a digraph, $D(G)$
\cite{graphtheory}.  We denote a directed (oriented) edge joining a vertex
$v_i$ to a vertex $v_j$, with the arrow pointing from $v_i$ to $v_j$, as ${\vec
  e}_{i,j}$, and the set of edges of $D(G)$ with such arrow assignments as
${\vec E}$, so that $D(G)=(V, {\vec E})$. For a given $G$, there are
\beq
N_{eo}(G) = 2^{e(G)}
\label{neo}
\eeq
such assignments of arrows to the edges of $G$ and hence $2^{e(G)}$
corresponding directed graphs $D(G)$. The subscript $eo$ on $N_{eo}(G)$ stands
for ``edge orientations''.  Since we study edge orientations of graphs $G$, we
assume henceforth that $G$ has at least one edge, and hence exclude the trivial
case in which $G$ consists only of one or more disjoint vertices with no
edges. Given a vertex $v_i$ in a digraph $D(G)$, its out-degree,
$\Delta^{+}(v_i)$, and in-degree, $\Delta^{-}(v_i)$, are, respectively, the
number of outgoing and incoming arrows on edges incident on $v_i$, so that the
total degree of the vertex is $\Delta(v_i) =
\Delta^{+}(v_i)+\Delta^{-}(v_i)$. A graph with the property that all vertices
have the same degree is denoted a $\Delta$-regular graph.  A directed cycle on
a directed graph $D(G)$ is defined as a set of directed edges forming a cycle
such that, as one traverses the cycle in a given direction, all of the arrows
on the oriented edges point in the direction of traversal. An acyclic
orientation of the edge arrows of $D(G)$ is one in which there are no directed
cycles. Further background on graph is given in the Appendix.

An important question in the study of directed graphs concerns the enumeration
of the subset of the $2^{e(G)}$ directed graphs $D(G)$ that are acyclic.  Since
this number depends only on the structure of $G$ itself, it is commonly denoted
$a(G)$ and is called the number of acyclic orientations of (arrows on edges of)
$G$. In addition to its intrinsic interest in mathematical graph theory, the
quantity $a(G)$ is also of interest in applications, such as manufacturing and
operations research.  The reason for this is that directed graphs that describe
these applications are acyclic.

The number of acyclic orientations of a graph $G$ can be calculated via an
evaluation of the chromatic polynomial of $G$, We recall that the chromatic
polynomial of a graph $G$, denoted $P(G,q)$, enumerates the the number of
assignments of $q$ colors to the vertices of $G$, subject to the condition that
no two adjacent vertices have the same color
\cite{graphtheory}-\cite{dkt}. This is called a proper $q$-coloring of (the
vertices of) $G$. $P(G,q)$ is a polynomial of degree $n$ in $q$. For proper $q$
colorings, $q$ must be a positive integer, and cannot be equal to 1 if $G$
contains at least one edge. More generally, one can consider the behavior of
$P(G,q)$ at other values of $q$.  For acyclic orientations, one has
\cite{stanley73} (see also \cite{winder66})
\beq
a(G) = (-1)^{n(G)} \, P(G,-1) \ .
\label{a_pqm1}
\eeq
The chromatic polynomial is a special case of an important two-variable
function, namely the partition function of the $q$-state Potts model
\cite{wurev} , $Z(G,q,v)$ with $v=-1$ (zero-temperature Potts antiferromagnet),
or equivalently, the Tutte polynomial $T(G,x,y)$ \cite{graphtheory,tutte54,
  tutte67,bo} with $x=1-q$ and $y=0$ (see Eqs. (\ref{pz}) and (\ref{zt}) in the
Appendix; some recent reviews are
\cite{welsh_merino2000,jemrev,jemcrev}). Using this connection, one can
equivalently express $a(G)$ as an evaluation of $T(G,x,y)$, namely
\beq
a(G) = T(G,2,0) \ .
\label{a_tx2y0}
\eeq

In many practical applications such as manufacturing processes and scheduling,
the relevant digraph is characterized by a single source vertex, e.g., the
first position of an item on an assembly line in a factory.  Thus, a second
quantity of interest can be defined as follows.  In a graph $G=(V,E)$, let us
pick a given vertex $v_i \in V$.  Among the $a(G)$ acyclic orientations of the
edges of $G$, count the number for which two conditions are satisfied: (i)
$v_i$ is a source vertex, i.e., it has only outgoing arrows on edges incident
with it (and hence maximal out-degree $\Delta^+(v_i)=\Delta(v_i) \ge 1$); and
(ii) $v_i$ is the only source vertex.  In order for this to be a function of
$G$, it must be true, and it does turn out to be true, that this number is
independent of which vertex $v_i$ one selects for this enumeration
\cite{greene_zaslavsky83}. This number is denoted as $a_0(G)$ and can also be
calculated from a knowledge of the chromatic polynomial $P(G,q)$.  This
polynomial $P(G,q)$ is identical to the partition function of the
zero-temperature $q$-state Potts antiferromagnet in statistical physics. A
proper $q$-coloring of $G$ is obviously not possible if $q=0$, so $P(G,0)=0$,
and hence, as a polynomial, $P(G,q)$ always has an overall factor of $q$.
Hence, one can define the reduced ($r$) polynomial $P_r(G,q)=q^{-1}P(G,q)$, as
in Eq. (\ref{pr}).  Then \cite{greene_zaslavsky83,gebhard_sagan99}
\beq
a_0(G) = (-1)^{n(G)-1} \, P_r(G,0) \ . 
\label{a0_prq0}
\eeq
From Eq. (\ref{pz}) and (\ref{zt}), $a_0(G)$ can also be defined as an
evaluation of the Tutte polynomial,
\beq
a_0(G) = T(G,1,0) \ .
\label{a0_tx1y0}
\eeq
If a graph $G$ contains a loop, this precludes the possibility of a proper
$q$-coloring, and thus the chromatic polynomial $P(G,q)$ vanishes identically,
as do both $a(G)$ and $a_0(G)$.  It is to avoid these trivial zeros that we
exclude graphs with loops in our analysis.

Let us illustrate these definitions in the simplest nontrivial case, namely the
tree graph $T_2$ with two vertices and a single edge joining them. In general,
for the tree graph with $n$ vertices, $T(T_n,x,y)=x^{n-1}$.  There are two edge
orientations of $T_2$, both of which are acyclic, so $a(T_2)=T(T_2,2,0)=2$. In
both of these, there is a source vertex, but the source vertex is different for
the two different edge orientations. Recalling the definition of $a_0(G)$, one
picks a specific vertex and then enumerates how many of the acyclic
orientations have this specific vertex as a source vertex, and this number is
$a_0(T_2)=T(T_2,1,0)=1$. As is obvious from the fact that the reversal of all
arrows is an automorphism of the set of all orientations of directed edges, one
could equivalently define $a_0$ as the number of acyclic orientations of $G$
such that there is a unique sink rather than a unique source.

Since acyclic orientations with a unique source are a subset of all acyclic
orientations, it follows that 
\beq
a_0(G) \le a(G) \ , \quad i.e., \quad T(G,1,0) \le T(G,2,0) \ .  
\label{aa0_inequality}
\eeq
This inequality is evident from Eq. (\ref{tij}), since the coefficients of the
nonzero terms in $T(G,x,y)$ are positive.  From the relations (\ref{a_tx2y0})
and (\ref{a0_tx1y0}), the necessary and sufficient condition for
(\ref{aa0_inequality}) to be an equality is clear, namely that 
$T(G,1,0) = T(G,2,0)$ if and only if $T(G,x,y)$ contains an overall factor of 
$y$ so that $T(G,1,0)=T(G,2,0)=0$ and thus $a_0(G)=a(G)=0$. 
For the lattice graphs of interest here, (\ref{aa0_inequality}) will be
a strict inequality, i.e., $a_0(G) < a(G)$. 

It is also of interest to enumerate, for a given graph $G$, the
number of digraphs $D(G)$ in which every directed edge is a
member of at least one directed cycle.  Such digraphs are called totally
cyclic, and the directed edges are called as totally cyclic orientations of
$D(G)$.  We denote the number of these as $b(G)$.  
(The number of totally cyclic orientations of $G$ should not be confused with
the number of linearly independent cycles on $G$, denoted $c(G)$, which is
given by $c(G) = e(G)+k(G)-n(G)$.) The number $b(G)$ can be obtained as
an evaluation of the Tutte polynomial, namely \cite{lasvergnas77,lasvergnas80}
\beq
b(G) = T(G,0,2) \ . 
\label{b_tx0y2}
\eeq
Starting with a given graph $G$, one can increase $b(G)$ arbitrarily by
replacing each edge with multiple edges joining the same pair of vertices.  In
order to have a minimal measure of totally cyclic orientations, we thus exclude
graphs with multiple edges in our analysis.  Using the equivalence between the
Tutte polynomial and the partition function of the Potts model, as given in
Eq. (\ref{zt}), we can express $b(G)$ as
\beq
b(G) = (-1)^{k(G)}Z(G,-1,1) = -Z(G,-1,1) \ . 
\label{bzqm1v1}
\eeq
where $Z(G,q,v)$ is the partition function of the Potts model on the graph $G$
at a temperature given by the variable $v$ defined in Eq. (\ref{veq}). 
Thus, $b(G)$ is obtained as an evaluation of the partition function of the
Potts ferromagnet at $q=-1$ and the (finite-temperature) value $v=1$. 
The quantity $b(G)$ can also be obtained as an evaluation of a one-variable
polynomial, namely the flow polynomial $F(G,q)$, as
\beq
b(G) = (-1)^{e(G)-n(G)-1}F(G,-1) \ . 
\label{bflow}
\eeq
The flow polynomial $F(G,q)$ enumerates the number of nowhere-zero $q$-flows on
the graph $G$ (with flow conservation mod $q$ at vertices) \cite{graphtheory}.
In addition to Refs. \cite{stanley73}-\cite{lasvergnas80}, some relevant
previous studies of these quantities $a(G)$, $a_0(G)$, and $b(G)$ include
Refs. \cite{merino_welsh99}-\cite{garijo2014}. In particular, in \cite{ka3}
we presented a number of results on acyclic orientations and their asymptotic
behavior.

A recursive family of graphs is a family of graphs such that the $(m+1)$'th
member, $G_{m+1}$, can be obtained from the $m$'th member, roughly speaking, by
the addition of some subgraph \cite{bds,biggsmer,biggscoloring}).  For
example, a square-lattice ladder strip of length $m+1$ vertices with free
boundary conditions can be obtained by adding a square to the end of the
square-lattice strip of length $m$. For a wide variety of recursive families of
graphs, these numbers $a(G)$, $a_0(G)$, and $b(G)$ grow exponentially with the
number of vertices, $n$, for $n >> 1$.  It is thus of interest to study the
associated exponential growth constants. Let us denote $\{ G \}$ as the limit
of the recursive family of $n$-vertex graphs $G$ as $n \to \infty$. We define
the three exponential growth constants for $a(G)$, $a_0(G)$, and $b(G)$ as
\beq
\alpha(\{ G \}) = \lim_{n \to \infty} [a(G)]^{1/n}
\label{alpha}
\eeq
\beq
\alpha_0(\{ G \}) = \lim_{n \to \infty} [a_0(G)]^{1/n}
\label{alpha0}
\eeq
and
\beq
\beta(\{ G \}) = \lim_{n \to \infty} [b(G)]^{1/n} \ . 
\label{beta}
\eeq
For all of the lattices that we study, we find the inequality 
\beq
\alpha_0(\{ G \}) < \alpha(\{G\}) \ . 
\label{alpha0_lt_alpha}
\eeq
Note that this inequality is not implied by the inequality 
(\ref{aa0_inequality}), since, {\it a priori}, the difference, 
$\lim_{n(G) \to \infty} [a(G)]^{1/n(G)} - 
 \lim_{n(G) \to \infty} [a_0(G)]^{1/n(G)}$ might vanish as $n(G) \to \infty$. 

Let $G$ be a planar graph, which we denote as $G_{pl}$, and denote $G_{pl}^*$
as the planar dual graph, formed by bijectively associating the vertices
(respectively faces) of $G_{pl}$ with the faces (respectively, vertices) of
$G_{pl}^*$, and connecting the vertices of $G_{pl}^*$ via edges crossing the
edges of $G_{pl}$.  For such a planar graph, the Tutte polynomial satisfies the
relation
\beq
T(G_{pl},x,y) = T(G_{pl}^*,y,x) \ . 
\label{tdual}
\eeq
In particular, $T(G_{pl},2,0) = T(G_{pl}^*,0,2)$, so 
\beq
a(G_{pl}) = b(G_{pl}^*) \ . 
\label{ab}
\eeq

We denote the number of faces of a graph $G$ as $f(G)$ and recall the Euler
relation that for a planar graph $G_{pl}$,
\beq
f(G_{pl})-e(G_{pl})+n(G_{pl})=2 \ . 
\label{euler}
\eeq
From the duality relation, it follows that $n(G_{pl}^*)=f(G_{pl})$. 
For $\Delta$-regular graphs $G$, 
\beq
e(G) = \frac{\Delta(G) \, n(G)}{2} \ .
\label{egdelta}
\eeq
For a $\Delta$-regular planar graph $G_{pl}$ we define the ratio
\beq
\nu_{ \{ G_{pl} \} } \equiv \lim_{n(G_{pl}) \to \infty} 
\frac{n(G_{pl}^*)}{n(G_{pl})} = \frac{\Delta(G_{pl})}{2}-1 \ . 
\label{nu_g}
\eeq
where we have used Eq. (\ref{egdelta}) in the last equality in (\ref{nu_g}). 
Note that 
\beq
\nu( \{ G_{pl} \}) = \frac{1}{ \nu( \{ G_{pl}^* \} ) } \ . 
\label{nunuinverse}
\eeq
We record the specific values 
\beq
\nu(sq) = 1 
\label{nusq}
\eeq
and 
\beq
\nu(tri) = \frac{1}{\nu(hc)} = 2 \ , 
\label{nuhctri}
\eeq
where the property that $\nu(tri) = 1/\nu(hc)$ follows from the fact that the
triangular and honeycomb lattice are planar duals of each other. 
From Eq. (\ref{ab}), it follows that if a planar graph is self-dual, indicated
as $G_{pl.,sd.}$, then $a(G_{pl.,sd.})=b(G_{pl.,sd.})$, and hence 
\beq
\alpha(\{G_{pl.,sd.}\}) = \beta(\{G_{pl.,sd.}\}) \ .
\label{alphabeta_rel}
\eeq
In particular, since the square lattice is planar and self-dual, we have 
\beq
\alpha(sq) = \beta(sq) \ , 
\label{alfbet_sq}
\eeq
so that the lower and upper bounds that we infer below for $\alpha(sq)$ also 
hold for $\beta(sq)$. For the triangular and honeycomb lattices, we obtain 
the relations
\beq
\alpha(hc) = [\beta(tri)]^{\nu(hc)} = [\beta(tri)]^{1/2}  
\label{betatri_alphahc}
\eeq
and
\beq
\beta(hc) = [\alpha(tri)]^{\nu(hc)} = [\alpha(tri)]^{1/2} \ . 
\label{betahc_alphatri}
\eeq

As was true of $a(G)$, $a_0(G)$, and $b(G)$, the corresponding exponential
growth constants $\alpha(\{ G \})$,  $\alpha_0(\{ G \})$, and 
$\beta(\{ G \})$ have interesting connections with quantities in statistical
physics. Specifically,
\beq
\alpha(\{G\}) = |W(\{G\},-1)|
\label{alpha_wqm1}
\eeq
and
\beq
\alpha_0(\{G\}) = |W(\{G\},0)| \ , 
\label{alpha0_wq0}
\eeq
where $W(\{G\},q)$ is the ground-state (i.e., zero-temperature) degeneracy of
states, normalized per vertex (site) of the $q$-state Potts antiferromagnet,
defined in Eq. (\ref{w}). The absolute values are used in Eqs.
(\ref{alpha_wqm1}) and (\ref{alpha0_wq0}) because for (real)
values of $q$ away from the positive integers, $P(G,q)$ can be negative, so
that the formal relation $W(\{G\},q)=\lim_{n \to \infty} [P(G,q)]^{1/n}$ in
Eq. (\ref{w}) requires specification of which of the $n$ roots of $(-1)$ one
chooses, while the value of $|W(\{G\},q)|$ is unambiguous. (Note that for any
finite $n=n(G)$, a negative sign in $P(G,q)$ is cancelled by the factor of
$(-1)^{n(G)}$ in Eq. (\ref{a_pqm1}) for $a(G)$ and the factor of
$(-1)^{n(G)-1}$ in Eq. (\ref{a0_prq0}) for $a_0(G)$.)

Furthermore,
\beq
\beta(\{G\}) = e^{|f(\{G\},-1,1)|} \ , 
\label{beta_ef}
\eeq
where $f(\{G\},q,v)$ is the dimensionless free energy per vertex of the
$q$-state Potts model defined in Eq. (\ref{f}).  Although the Potts model
partition function is naturally defined for integral $q \ge 1$ in statistical
mechanics, its definition can also be extended, via Eq. (\ref{zcluster}) to
more general values of $q$, and this generalization is used here. The absolute
value is used in Eq. (\ref{beta_ef}) because $Z(G,-1,1)$ is negative (as is
clear from Eq. (\ref{bzqm1v1})), so the formal relation (\ref{f}) requires
specification of which of the $n$ roots of $(-1)$ one uses, whereas
$|f(\{G\},-1,1)|$ is unambiguously determined by Eq. (\ref{f}).

Thus, although it might initially seem that $a(G)$, $a_0(G)$, $b(G)$, and
the associated exponential growth constants $\alpha(\{G\})$, $\alpha_0(\{G\})$,
and $\beta(\{G\})$, are only of relevance in mathematical graph theory and
applications such as operations research, Eqs. (\ref{a_pqm1})-(\ref{b_tx0y2})
and (\ref{alpha_wqm1}), (\ref{alpha0_wq0}), and (\ref{beta_ef}), with 
Eqs. (\ref{pz}), (\ref{f}), (\ref{w}), and (\ref{zt}), show
that these quantities also have interesting and fruitful connections with
statistical physics. Our new results in this paper demonstrate the value of
exploiting these connections.

A basic property of a digraph is that the number of edge orientations on $G$,
$N_{eo}(G)$, grows exponentially rapidly as a function of the number of edges
of $G$.  In order to define the corresponding exponential growth constant, one
first expresses $e(G)$ in terms of $n(G)$. This is done via Eq.
(\ref{egdelta}) for $\Delta$-regular graphs.  More generally, for graphs
containing vertices of different degrees, we define an effective vertex degree,
as in \cite{wn} (with $n \equiv n(G)$ here and below),
\beq
\Delta_{eff}(G) = \frac{2e(G)}{n} \ . 
\label{deffgn}
\eeq
In particular, in the $n \to \infty$ limit, 
\beq
\Delta_{eff}( \{ G \}) = \lim_{n(G) \to \infty} \frac{2e(G)}{n} \ . 
\label{deff}
\eeq
Thus, the exponential growth constant for the total number of edge orientations
$D(G)$ of a $\Delta$-regular family of graphs $G$, normalized per vertex, is
\beq
\epsilon(\{ G \}) \equiv \lim_{n \to \infty} [N_{eo}(G)]^{1/n} = 
\lim_{n \to \infty} (2^{e(G)})^{1/n} = 2^{\Delta(\{ G \})/2} \ . 
\label{epsilon}
\eeq
More generally, for a family of graphs $G$ containing vertices with
different degrees, 
\beq
\epsilon(\{ G \}) = 2^{\Delta_{eff}(\{ G \})/2} \ . 
\label{epsilon_eff}
\eeq

For a given digraph, many of the edge orientations do not fall into any of the
three classes (i)-(iii), i.e., they are not acyclic or totally cyclic.  This is
reflected in the property that for most digraphs, $a(G)+b(G) < N_{eo}(G)$.  An
interesting question related to this is the following: for a given graph
$G=G(V,E)$, what fraction of the total number of edge orientations
is comprised of those that are (i) acyclic, (ii) acyclic
with a unique source vertex, and (iii) totally cyclic?  Each of the
corresponding numbers $a(G)$, $a_0(G)$, and $b(G)$ can be denoted generically
as $N_{cond.}(G)$, where the subscript $cond.$ refers to the condition
that the edge orientations must satisfy to be a member of the
given class.  The corresponding fraction is then the ratio
\beq
r_{cond.}(G) \equiv \frac{N_{cond.}(G)}{N_{eo}(G)} 
= \frac{N_{cond.}(G)}{2^{e(G)}} \ . 
\label{rcond}
\eeq

For each of the specified conditions (i)-(iii), $r_{cond.}(G) \le 1$.  For
certain families of graphs, $r_{cond.}(G) = 1$. Recall that a tree graph 
is a connected graph that does not contain any circuits. For $n$-vertex
tree graphs, denoted $T_n$, all edge orientations are acyclic, so
$a(T_n)=N_{eo}(T_n)=2^{n-1}$ and thus $r_a(T_n)=1$, while $b(T_n)=0$, so
$r_b(T_n)=0$ (independent of $n$).  However, as discussed above, from exact
results on strip graphs of lattices with fixed width $L_y \ge 2$ and with
various transverse and longitudinal boundary conditions, one finds that
$r_{cond.}(G) < 1$ and, furthermore, that $N_{cond.}(G)$ grows
exponentially rapidly with $n(G)$ for $n(G) >> 1$.  A relevant question is the
following: for a given condition, does the ratio $r_{cond.}(G)$ approach a
finite nonzero constant as $n(G) \to \infty$ or not?  The answer to this
question depends on the type of families graphs that one considers.  If one
were to consider tree graphs, for example, then the ratio $r_a(T_n)$ would be
finite (and equal to its maximal value, 1) for all $n$ and, in particular, for
$n \to \infty$.  However, for all of the lattice strip graphs of finite width
(i.e., excluding the circuit graph) that we have studied, $r_{cond.}(G)$
vanishes as $n(G) \to \infty$.  That is,
\beq
\lim_{n \to \infty} \frac{N_{cond.}(G)}{N_{eo}(G)} = 0 \ , \quad {\rm where} \ 
N_{cond.}(G) = a(G), \ a_0(G), \ {\rm or} \  b(G) \ . 
\label{ncondlimt}
\eeq
Specifically, for these lattice strip graphs we find that $r_{cond.}(G)$
vanishes exponentially rapidly as $n(G) \to \infty$.  Therefore, one is
motivated to define a measure of this exponential decrease in the ratio
$r_{cond.}(G)$.  Since for most lattice strip graphs both the numerator and
denominator of the ratio $r_{cond.}(G)$ increase exponentially rapidly with
$n$, it is natural to define this measure as
\beq
\rho_{N_{cond.}}(\{G\})  \equiv \lim_{n \to \infty} [r_{cond.}(G)]^{1/n} 
      = \frac{\lim_{n \to \infty} [N_{cond.}(G)]^{1/n}}{\epsilon(\{G\})} \ . 
\label{rho}
\eeq
For each of the three quantities considered here corresponding to the
orientations satisfying the specified conditions (i) acyclic, (ii) acyclic with
a unique source vertex or sink vertex, and (iii) totally cyclic, we then have
\beq
\rho_\alpha(\{G\}) \equiv \lim_{n \to \infty} \bigg ( \frac{a(G)}{N_{eo}(G)} 
\bigg )^{1/n} = \frac{\alpha(\{ G \})}{\epsilon(\{ G \} )} 
\label{rho_alpha}
\eeq
\beq
\rho_{\alpha_0}(\{G\}) \equiv \lim_{n \to \infty} \bigg ( \frac{a_0(G)}
{N_{eo}(G)} \bigg )^{1/n} = \frac{\alpha_0(\{ G \})}{\epsilon(\{ G \} )} 
\label{rho_alpha0}
\eeq
and
\beq
\rho_\beta(\{G\}) \equiv \lim_{n \to \infty} \bigg ( \frac{b(G)}{N_{eo}(G)} 
\bigg )^{1/n} = \frac{\beta(\{ G \})}{\epsilon(\{ G \} )} \ . 
\label{rho_beta}
\eeq

This paper is organized as follows. In Section \ref{example_section} we present
a number of exact results on $a(G)$, $a_0(G)$, $b(G)$ for lattice strip graphs
and show how, in the limit of infinite strip length, these yield resultant
values for the corresponding exponential growth constants $\alpha(G)$,
$\alpha_0(G)$, and $\beta(G)$. In Sections \ref{stripcal_section} and
\ref{method_section} we discuss our methods for inferring lower and upper
bounds on the exponential growth constants for infinite lattices from
calculations on strip graphs of varying widths in the limit of infinite width.
In Sections \ref{alfalf0_values_section} and \ref{beta_values_section} we
present our numerical results on these lower and upper bounds for
$\alpha(\Lambda)$, $\alpha_0(\Lambda)$, and $\beta(\Lambda)$ for various
lattices $\Lambda$. In these sections, using exact values of $\alpha(tri)$,
$\alpha_0(tri)$, and $\beta(hc)$, we show that our lower and upper bounds on
these quantities are very close to the exact values for modest values of strip
widths.  We present some further discussion in Section
\ref{comparison_section}, including a comparison with growth constants for
spanning trees.  Our conclusions are given in Section
\ref{conclusion_section}.  Some graph theory background is included in Appendix
\ref{graphtheory}.


\section{Exact Results for Lattice Strip Graphs}
\label{example_section}

In this section we present some exact calculations of $a(G)$, $a_0(G)$, and
$b(G)$ for lattice strip graphs of fixed transverse width $L_y$ and arbitrarily
great length $L_x$ with certain boundary conditions, and show how one derives
the corresponding exponential growth constants $\alpha(\{G\})$,
$\alpha_0(\{G\})$, and $\beta(\{G\})$ from these in the limit $L_x >> 1$.  As
indicated, we take the longitudinal direction to lie along the $x$ (horizontal)
axis, and the transverse direction to lie along the $y$ (vertical) axis. We
also include results on the constants $\rho_\alpha(\{G\})$,
$\rho_{\alpha_0}(\{G\})$, and $\rho_\beta(\{G\})$.  These examples are selected
from calculations of chromatic and Tutte polynomials for a number of lattice
strip graphs (e.g., \cite{ka3}, \cite{w}-\cite{zttor}).


\subsection{Cyclic Square-Lattice Ladder Graph}

We first consider the square-lattice ladder strip graph $L_m$ of width $L_y=2$
vertices and length $L_x \equiv m$ vertices with cyclic longitudinal boundary
conditions and free transverse boundary conditions.  This graph has $n(L_m)=2m$
vertices, $e(L_m)=3m$ edges, and uniform vertex degree, $\Delta=3$. The 
number of linearly independent cycles on $L_m$ is $c(L_m)=m+1$. We denote the
infinite-length limit, $m \to \infty$, of this strip graph as $\{ L \}$. The
exponential growth constant for the number of edges in this limit is
\beq
\epsilon(\{ L \} ) = 2^{3/2} = 2.828427 \ ,
\label{eps_sqlad}
\eeq
where here and below, we write non-integer numbers with the indicated number of
significant figures.

Evaluating the chromatic polynomial at $q=-1$, one finds \cite{ka3}
\beq
a(L_m) = 7^m - 2(2^m+4^m)+5 \ . 
\label{a_sqlad}
\eeq
Using Eq. (\ref{alpha}), one calculates 
\beq
\alpha(\{L\})=\sqrt{7} = 2.645751 \ .
\label{alpha_sqlad}
\eeq

Evaluating the expression for $a_0(G)$ in Eq. (\ref{a0_prq0}), we obtain 
\beq
a_0(L_m) = (2m-3) \, 3^{m-1} - (m-2) \ . 
\label{a0_sqlad}
\eeq
Hence, in the limit $m \to \infty$, 
\beq
\alpha_0(\{ L \}) = \sqrt{3} = 1.732051 \ . 
\label{alpha0_sqlad}
\eeq
The origin of the factors of $m$ in Eq. (\ref{a0_sqlad}) is as follows.
For a cyclic strip graph $G_{strip}$, the chromatic polynomial $P(G_{strip},q)$
has the form of a sum of powers of certain algebraic functions multiplied by
various coefficients, given in Eq. (\ref{pcyc}) in the Appendix.  These powers
involve $L_x=m$, the length of the strip. Although a chromatic polynomial
always has a factor of $q$, this factor is not explicit in the expression
written as a sum of powers of these algebraic functions.  Consequently, to
evaluate $P_r(G_{strip},q)=q^{-1}P(G_{strip},q)$ at $q=0$, one actually uses
L'H\^{o}pital's rule, calculating
\beq
P_r(G_{strip},0) = \lim_{q \to 0} P_r(G_{strip},q) =
\frac{dP(G_{strip},q)}{dq}\Big |_{q=0}
\label{prcalc}
\eeq
It is this differentiation that brings down factors of $m$. 

For the number of totally cyclic orientations of $G$, from the solution for
$Z(L_m,q,v)$ or equivalently $T(L_m,x,y)$ in Ref. \cite{a}, we calculate 
\beq
b(L_m) = 2(4^m)-3^m-5 \ , 
\label{b_sqlad}
\eeq
so that, in the limit $m \to \infty$, 
\beq
\beta( \{ L \}) = 2 \ .
\label{beta_sqlad}
\eeq

From these results and the expression for $\epsilon(\{L\})$, we compute
\beq
\rho_\alpha(\{ L \}) = \frac{\alpha(\{ L \})}{\epsilon(\{L\})} = 
\sqrt{\frac{7}{8}} = 0.935414
\label{rho_alpha_sqlad}
\eeq
\beq
\rho_{\alpha_0}(\{ L \}) = \frac{\alpha_0(\{ L \})}{\epsilon(\{L\})} = 
\sqrt{\frac{3}{8}} = 0.612372
\label{rho_alpha0_sqlad}
\eeq
and
\beq
\rho_\beta(\{ L \}) = \frac{\beta(\{ L \})}{\epsilon(\{L\})} = 
\frac{1}{\sqrt{2}} = 0.707107 \ . 
\label{rho_beta_sqlad}
\eeq
%


\subsection{Cyclic Triangular-Lattice Ladder Graph}

We next consider a cyclic strip of the triangular lattice $TL_m$ [where
$TL$ is an abbreviation for ``triangular (lattice) ladder''] of width
$L_y=2$ vertices and length $L_x=m$ vertices.  This graph can be obtained from
the cyclic square-lattice strip by adding a diagonal edge to each square from,
say, the lower left vertex to the upper right vertex.  This graph has
$n(TL_m)=2m$ vertices, $e(L_m)=4m$ edges, uniform vertex degree
$\Delta=4$, and $c(L_m)=2m+1$ linearly independent cycles. Evaluating the
chromatic polynomial at $q=-1$, one finds \cite{ka3}
\beq
a(TL_m) = 9^m - 2\Bigg [ \bigg ( \frac{7+\sqrt{13}}{2} \  \bigg )^m 
+ \bigg ( \frac{7-\sqrt{13}}{2} \  \bigg )^m \Bigg ] + 5 \ . 
\label{a_trilad}
\eeq
Denoting the limit of $TL_m$ as $m \to \infty$ as $\{ TL \}$ and using 
Eq. (\ref{alpha}), one has
\beq
\alpha(\{TL\})=3 \ . 
\label{alpha_trilad}
\eeq
Evaluating $P_r(TL_m,q)$ at $q=0$, we obtain 
\beq
a_0(TL_m) = \Big ( \frac{2m}{3}-1 \Big ) 4^m - \frac{2m}{3} + 2 \ ,
\label{asource_trilad}
\eeq
and hence 
\beq
\alpha_0(\{TL \}) = 2 \ . 
\label{alpha0_trilad}
\eeq
For the number of totally cyclic orientations of $G$, from the solution for
$Z(TL_m,q,v)$ in Ref. \cite{ta}, we find 
\beq
b(TL_m) = 2\Bigg [ \bigg ( \frac{11+3\sqrt{13}}{2} \bigg )^m + 
  \bigg ( \frac{11-3\sqrt{13}}{2} \bigg )^m \Bigg ] - 9^m - 5 \ , 
\label{b_trilad}
\eeq
and hence 
\beq
\beta( \{ TL \}) = \Bigg ( \frac{11+3\sqrt{13}}{2} \Bigg )^{1/2} = 
\frac{3+\sqrt{13}}{2} = 3.3027756 \ . 
\label{beta_trilad}
\eeq

Combining these results with 
\beq
\epsilon(\{ TL \}) = 4  \ , 
\label{eps_trilad}
\eeq
we obtain
\beq
\rho_\alpha(\{ TL \}) = \frac{3}{4} \ , 
\label{rho_alpha_trilad}
\eeq
\beq
\rho_{\alpha_0}(\{ TL \}) = \frac{1}{2} \ , 
\label{rho_alpha0_trilad}
\eeq
and
\beq
\rho_\beta(\{ TL \}) = \frac{3+\sqrt{13}}{8} = 0.825694 \ . 
\label{rho_beta_trilad}
\eeq
%


\subsection{Cyclic Honeycomb-Lattice Ladder Graph}

We next consider a cyclic strip of the honeycomb lattice, $HL_m$ [where $HL$
stands for ``honeycomb (lattice) ladder''] of width $L_y=2$ vertices and length
$L_x=2m$ vertices.  Here, $m$ is the number of hexagons in a horizontal layer
of the strip. This graph can be obtained from the cyclic square-lattice strip
by adding a vertex on each horizontal edge (which is a homeomorphic expansion
of the square-lattice ladder strip).  The graph $HL_m$ has $n(HL_m)=4m$
vertices, $e(L_m)=5m$ edges, and $c(L_m)=m+1$ linearly independent cycles. It
has two equal subsets of vertices of two different degree values, namely 2 and
3, and thus an effective vertex degree of $\Delta_{eff}=5/2$. Evaluating the
chromatic polynomial \cite{pg} at $q=-1$, we find
\beq
a(HL_m) = (31)^m - 2( 10^m + 4^m ) + 5 \ . 
\label{a_hclad}
\eeq
Denoting the limit of $HL_m$ as $m \to \infty$ as $\{ HL \}$ and using 
Eq. (\ref{alpha}), we thus obtain
\beq
\alpha(\{HL\})=(31)^{1/4} = 2.359611 \ . 
\label{alpha_hclad}
\eeq
Further, we compute 
\beq
a_0(HL_m) = \bigg ( \frac{6m}{5}-1 \bigg ) \, 5^m -2(m-1) \ , 
\label{asource_hclad}
\eeq
and hence 
\beq
\alpha_0(\{HL \}) = 5^{1/4} = 1.495349 \ . 
\label{alpha0_hclad}
\eeq
For the number of totally cyclic orientations of $G$, from the solution for
$Z(HL_m,q,v)$ or equivalently $T(HL_m,x,y)$ in Ref. \cite{hca}, we calculate
\beq
b(HL_m) = 2 \, (4^m) - 3^m - 5 \ , 
\label{b_hclad}
\eeq
and hence 
\beq
\beta( \{ HL \}) = \sqrt{2} = 1.414214 \ . 
\label{beta_hclad}
\eeq
Combining these results with 
\beq
\epsilon( \{ HL \}) = 2^{5/4} = 2.378414 \ , 
\label{eps_hclad}
\eeq
we obtain 
\beq
\rho_\alpha(\{ HL \}) = \Big ( \frac{31}{32} \Big )^{1/4} = 0.992094
\label{rho_alpha_hclad}
\eeq
\beq
\rho_{\alpha_0}(\{ HL \}) = \Big ( \frac{5}{32} \Big )^{1/4} = 0.628717 
\label{rho_alpha0_hclad}
\eeq
and
\beq
\rho_\beta(\{ HL \}) = \Big ( \frac{1}{8} \Big )^{1/4} = 0.594604 \ . 
\label{rho_beta_hclad}
\eeq
%


\subsection{A Family of Self-Dual Planar Graphs}

An $n$-vertex wheel graph $Wh_n = K_1 + C_{n-1}$ is comprised of a circuit
graph with all $n-1$ vertices on the ``rim'' connected to one central ``spoke''
vertex. (Here, $G+H$ is the ``join'' of $G$ and $H$).  This graph has
$e(Wh_n)=2(n-1)$ and thus $c(Wh_n)=n-1$ linearly independent circuits.  The
$n-1$ vertices on the rim have degree 3 and the central spoke vertex has degree
$n-1$, so in the limit $n \to \infty$, the effective degree is
$\Delta_{eff}=4$. The wheel graph $Wh_n$ is a self-dual planar graph, so, as a
consequence of Eq. (\ref{ab}),
\beq
a(Wh_n)=b(Wh_n) \ . 
\label{abwheel}
\eeq
An elementary calculation yields 
$P(Wh_n,q) = q[(q-2)^{n-1}+(q-2)(-1)^{n-1}]$, so 
\beq
a(Wh_n) = b(Wh_n) = 3^{n-1}-3
\label{a_wheel}
\eeq
and
\beq
a_0(Wh_n)=2^{n-1}-2 \ . 
\label{a0_wheel}
\eeq
Denoting $\{Wh\}$ as the $n \to \infty$ limit of the $Wh_n$ family, we then
have 
\beq
\alpha(\{Wh \}) = \beta(\{ Wh \}) = 3
\label{alphabeta_wheel}
\eeq
and
\beq
\alpha_0(\{Wh\})=2 \ . 
\label{alpha0_wheel}
\eeq
We note that $Wh_n$ (and its $n \to \infty$ limit) share with the infinite
square lattice the property of being planar and self-dual. 
Combining the results above with 
\beq
\epsilon( \{ Wh \} ) = 4 \ , 
\label{eps_wheel}
\eeq
we find 
\beq
\rho_\alpha(\{ Wh \}) = \rho_\beta(\{ Wh \}) = \frac{3}{4} 
\label{rho_alphabeta_wheel}
\eeq
and
\beq
\rho_{\alpha_0}(\{ Wh \}) = \frac{1}{2} \ . 
\label{rho_alpha_0_wheel}
\eeq
%


\subsection{The $sq_d$ Family of Cyclic Strip Graphs}

It is also of interest to investigate a family of strip graphs with a higher
value of $\Delta$.  We define strip graphs of this family, denoted $(sq_d)_m$
(where the subscript $d$ stands for ``diagonals''), as follows. One starts
with the cyclic square-lattice ladder graph $L_m$ of width $L_y=2$ and length
$L_x=m$ and then adds (i) an edge connecting the upper left and lower right
vertices of each square to each other, and (ii) an edge connecting the upper
right and lower left vertices of each square to each other. This is a
$\Delta$-regular nonplanar lattice graph with $\Delta_{sq_d}=5$. The cyclic
$(sq_d)_m$ lattice strip has $n=2m$ vertices and $5m$ edges. Using our
calculations in \cite{k}, we obtain
\beq
a([sq_d]_m)= (12)^m-2 \cdot (8)^m + 2^{m+1} 
\label{a_sqdm}
\eeq
and
\beq
a_0([sq_d]_m) = \Big ( \frac{m}{2}-1 \Big ) \, 6^m + 3 \cdot 2^{m-1} 
\label{a0_sqdm}
\eeq
Denoting $\{ sq_d \}$ as the $m \to \infty$ limit of the $[sq_d]_m$ 
family, we then have 
\beq
\alpha(\{ sq_d \}) = 2\sqrt{3} = 3.464102
\label{alpha_sqd_strip}
\eeq
and
\beq
\alpha_0(\{ sq_d \})= \sqrt{6} = 2.449490 \ . 
\label{alpha0_sqd_strip}
\eeq
From our results in \cite{ka}, we find
\beqs
b([sq_d]_m) &=& 2 \Big [ (13+\sqrt{181} \, )^m + (13-\sqrt{181} \, )^m \Big ] 
- \Big [ \{ 2(6+\sqrt{39} \,) \}^m + \{ 2(6-\sqrt{39} \, ) \}^m \Big ] 
-2^{m+1} \ , 
\cr\cr
&& 
\label{b_sqdm}
\eeqs
so that, in the $m \to \infty$ limit, 
\beq
\beta(\{ sq_d \}) = \sqrt{13 + \sqrt{181}} = 5.143309 \ . 
\label{beta_sqd_strip}
\eeq
In conjunction with 
\beq
\epsilon( \{ sq_d \} ) = 2^{5/2} = 5.656854 \ , 
\label{eps_sqd}
\eeq
these results yield 
\beq
\rho_\alpha(\{ sq_d \}) = \sqrt{\frac{3}{8}} = 0.612372
\label{rho_alpha_sqd}
\eeq
\beq
\rho_{\alpha_0}(\{ sq_d \}) = \frac{\sqrt{3}}{4} = 0.433013
\label{rho_alpha_0_sqd}
\eeq
and
\beq
\rho_{\beta_0}(\{ sq_d \}) = \sqrt{\frac{13+\sqrt{181}}{32}} = 0.909217 \ . 
\label{rho_beta_sqd}
\eeq
%


\subsection{Comparative Properties}

We list our results for the exponential growth constants and the corresponding
$\rho_{cond.}$ constants for these infinite-length, finite-width strips in
Table \ref{strip_table}. From these results, we can observe several properties.
We find that $\alpha(\{G\})$, $\alpha_0(\{G\})$, and $\beta(\{G\})$ 
are monotonically increasing functions
of $\Delta$ (and, where applicable, of $\Delta_{eff}$). This is also true of
the ratios $\alpha_0(\{ G \})/\alpha( \{ G \})$ for the various strips; 
from the values listed in Table \ref{strip_table}, we have
\beq
\frac{\alpha_0(\{ HL \})}{\alpha( \{ HL \})} = 
\bigg ( \frac{5}{31} \bigg )^{1/4} = 0.633727
\label{alpha0_over_alpha_hcstrip}
\eeq
\beq
\frac{\alpha_0(\{ L \})}{\alpha( \{ L \})} = \sqrt{\frac{3}{7}} = 0.654654
\label{alpha0_over_alpha_sqstrip}
\eeq
\beq
\frac{\alpha_0(\{ Wh \})}{\alpha( \{ Wh \})} = 
\frac{\alpha_0(\{ TL \})}{\alpha( \{ TL \})} = \frac{2}{3} 
\label{alpha0_over_alpha_whtristrip}
\eeq
and
\beq
\frac{\alpha_0(\{ sq_d \})}{\alpha( \{ sq_d \})} = \frac{1}{\sqrt{2}} =
  0.707107 \ . 
\label{alpha0_over_alpha_sqdtrip}
\eeq
Concerning the ratios $\rho_{EGC}(\{ G \})$, we find that 
$\rho_\alpha(\{G\})$ and $\rho_{\alpha_0}(\{G\})$ are monotonically 
decreasing functions, while $\rho_\beta(\{ G \})$ is a monotonically increasing
function of $\Delta$ (and, where applicable, of $\Delta_{eff}$). 


\subsection{Some Families of Graphs Without Exponential Growth for 
$a(G)$, $a_0(G)$, and/or $b(G)$}

For perspective, we mention some families of graphs $G$ for which
$a(G)$, $a_0(G)$, and/or $b(G)$ do not exhibit exponential growth with $n$.  
We begin with two recursive families of graphs, namely tree graphs and circuit
graphs.  For $n$-vertex tree graphs, $T_n$, 
two of the three quantities of interest here are actually constants,
independent of $n$.  These are $a_0(T_n)=1$, and $b(T_n)=0$. The other
quantity, $a(T_n)$ does grow exponentially with $n$ and, indeed, is maximal,
namely $a(T_n)=N_{eo}(T_n)=2^{n-1}$.  Hence, denoting $\{ T \}$ as the $n \to
\infty$ limit of the $T_n$ family, we have $\alpha( \{ T \}) = 2$, $\alpha_0(
\{ T \}) = 1$, and $\beta( \{ T \}) = 0$.  

For a circuit graph, $a(C_n)=2^n-2$,
$a_0(C_n)=n-1$, and $b(C_n)=2$, so that although $a_0(C_n)$ grows with $n$, it
does not grow exponentially rapidly, and $b(C_n)$ is a constant, independent of
$n$.  Hence, denoting $\{ C \}$ as the $n \to \infty$ limit of the $C_n$
family, it follows that $\alpha(\{ C \}) = 2$, $\alpha_0(\{ C \})=1$, and
$\beta(\{ C \})=1$.  In contrast, for strip graphs of various lattices with
widths $L_y \ge 2$, we do find that $a(G)$, $a_0(G)$, and $b(G)$ grow
exponentially with $n$. 

More generally, from a physics point of view, the property that $Z(G,q,v)$
grows exponentially rapidly with $n(G)$ as $n(G) \to \infty$ is equivalent to
the property that the free energy per vertex (or per $d$-dimensional volume,
for a $d$-dimensional lattice) is a constant in this limit.  In turn, this
reflects the extensivity of the total free energy in statistical physics.
However, even in this physics context, there are examples where $Z(G,q,v)$ does
not grow exponentially rapidly in the large-$n(G)$ limit.  We recall that the
chromatic polynomial is equal to the zero-temperature partition function of the
Potts antiferromagnet, Eq. (\ref{pz}). For a bipartite graph $G_{bip.}$, such
as the square or honeycomb lattices, $P(G_{bip.},2)=2$, independent of
$n(G_{bip.})$, while for a tripartite graph $G_{trip.}$, such as the triangular
lattice, $P(G_{trip.},3)=3!$, independent of $n(G_{trip.})$.  In both of these
cases, the chromatic polynomials evaluated at these respective values of $q$ do
not exhibit exponential growth with the number of vertices and, indeed, are
independent of the number of vertices.

It can also happen that for a family of graphs $G$, the quantities $a(G)$,
$a_0(G)$, and $b(G)$ grow more rapidly than exponentially with $n$. An example,
which is not a recursive family, is the family of complete graphs,
$K_n$. Recall that a complete graph $K_n$ is defined as a graph with $n$
vertices such that each vertex is connected to every other vertex by an edge,
so that there are $e(K_n) = {n \choose 2}$ edges. The chromatic polynomial for
$K_n$ is $P(K_n,q)=\prod_{j=0}^{n-1} (q-j)$. Therefore, from
Eqs. (\ref{a_pqm1}) and (\ref{a0_prq0}), one has
\beq
a(K_n) = n!
\label{akn}
\eeq
and
\beq
a_0(K_n) = (n-1)!
\label{a0kn}
\eeq
Since $n!$ has the asymptotic behavior given by $n! \sim (n/e)^n \, (2\pi
n)^{1/2}$ for $n >> 1$ (the Stirling formula), it follows that both $a(K_n)$
and $a_0(K_n)$ grow more rapidly than exponentially as $n \to \infty$. Having
mentioned these families of graphs for contrast, we return in the next section
to our main subject, namely strip graphs of various lattices.


\section{Calculations of $\alpha(\{G \})$ for Infinite-Length, Finite-Width 
Lattice Strips}
\label{stripcal_section}

\subsection{General} 

In this section we present calculations of $\alpha(\{ G \})$ for strip graphs
of several lattices, denoted generically as $\Lambda$ in the limit of infinite
length, $L_x \to \infty$, for various values of the width, $L_y$, and various
boundary conditions.  The results for $\alpha(\{ G \})$ and $\alpha_0(\{G \})$
for finite $L_y$ are independent of the boundary condition (free, periodic, or
twisted periodic) in the longitudinal ($x$) direction, denoted $BC_x$, but do
depend on the boundary condition in the transverse ($y$) direction, denoted
$BC_y$.  In detail, the relevant boundary conditions $(BC_y,BC_x)$ and their
names are as follows: (i) (F,F), free, (ii) (F,P), cyclic, (iii) (P,F),
cylindrical, (iv) (P,P), toroidal. In past work we have also considered (v)
(F,TP), M\"obius, and (vi) (P,TP), Klein-bottle, where here the symbol T stands
for twisted, but since strips with twisted longitudinal boundary conditions
yield the same results relevant here as the corresponding cyclic and toroidal
strips, it is not necessary to consider these twisted longitudinal boundary
conditions here.  For our present discussion, the infinite-length strip of a
lattice $\Lambda$ is indicated as $\{\Lambda,(L_y)_F \times \infty\}$ and
$\{\Lambda,(L_y)_P \times \infty\}$ for these two respective transverse
boundary conditions (with brackets included here for clarity).  For our
discussion of $\alpha$ and $\alpha_0$ on infinite-length lattice strips, we
will sometimes use an equivalent notation $\{\Lambda,(L_y) \times \infty,
free\}$ and $\{\Lambda,(L_y) \times \infty, cyl \}$, where the abbreviation
$cyl$ stands for cylindrical.

We will make use of a property of the chromatic and Tutte polynomials for these
strip graphs, namely that they can be written as a sum of certain coefficients
multiplied by powers of various functions, generically denoted $\lambda$,
depending on $\Lambda$, $L_y$, and the boundary conditions, but not on $L_x$.
The powers to which these $\lambda$ functions are raised are given by the
length, $m$, of the strip (see Appendix). As $m \to \infty$, the $\lambda$
function with the largest magnitude dominates the sum.  From calculations
of chromatic polynomials for strip graphs of various lattices
\cite{ka3}-\cite{zttor}, we know what this dominant $\lambda$ function is.  We
remark that for the strips that we consider, the dominant $\lambda$ function in
$P(G,q)$ at $q=-1$ and $q=0$ is the same as the $\lambda$ function that is
dominant at large $q$, and has coefficient $c^{(0)}=1$ in the notation of
Eqs. (\ref{pcyc}) and (\ref{cd}). To calculate $\alpha(\Lambda)$ and
$\alpha_0(\Lambda)$, we only need the dominant $\lambda$ function for the given
strip with free or periodic transverse boundary conditions, which will be
denoted as $\lambda_{\Lambda,L_y,free}(q)$ or $\lambda_{\Lambda,L_y,cyl}(q)$,
respectively. The reason that our results for $\alpha(\{ G \})$ and
$\alpha_0(\{ G \})$ are independent of the longitudinal boundary conditions is
that, as discussed in our earlier work, the dominant $\lambda$ for these is the
same for free and periodic (and twisted periodic) longitudinal boundary
conditions. In contrast, our results for $\beta(\{G \})$, to be discussed in
Section \ref{beta_values_section}, do depend on both the longitudinal and
transverse boundary conditions.

The resultant exponential growth constants for acyclic orientations and 
acyclic orientations with a unique source on the
infinite-length limits of the square and triangular lattice strips (which have
$n=L_xL_y$) are
\beq
\alpha(\Lambda,(L_y)_{BC_y} \times \infty) = \lim_{n \to \infty} 
[a(\Lambda,(L_y)_{BC_y} \times L_x)]^{1/n} = 
[\lambda(\Lambda,L_y,BC_y)(-1)]^{1/_{L_y}}
\label{alpha_infstrip}
\eeq
and
\beq
\alpha_0(\Lambda,(L_y)_{BC_y} \times \infty) = \lim_{n \to \infty} 
[a_0(\Lambda,(L_y)_{BC_y} \times L_x)]^{1/n} = 
[\lambda(\Lambda,L_y,BC_y)(0)]^{1/_{L_y}}
\label{alpha0_infstrip}
\eeq
For strips of the honeycomb lattice, $n=2mL_y$, so one replaces the 
exponents $1/L_y$ by $1/(2L_y)$ in Eqs. (\ref{alpha_infstrip}) and 
(\ref{alpha0_infstrip}). 

In our previous study \cite{ka3}, we showed that the resultant values of
$\alpha(\Lambda,(L_y)_F \times \infty)$ and $\alpha(\Lambda,(L_y)_P \times
\infty)$ were monotonically increasing functions of $L_y$ for the full range of
widths $L_y$ for which we carried out calculations with the square-lattice and
triangular-lattice strips. To anticipate the new results that we present here,
we continue to find this behavior both for these lattice strips and for all of
the other lattice strips that we have studied.  This provides strong additional
support for the inference that we made in Ref. \cite{ka3}, that for a given
infinite-length, finite-width strip of a lattice $\Lambda$ with free or
periodic transverse boundary conditions,
\beq 
\alpha(\Lambda,(L_y)_{BC_y} \times \infty) \ \ {\rm is \ a \
  monotonically \ increasing \ function \ of} \ L_y. 
\label{alpha_monotonic}
\eeq
Furthermore, our results in \cite{ka3} provide strong additional support for
the inference that
\beq
\lim_{L_y \to \infty} \alpha(\Lambda,(L_y)_{BC_y} \times \infty) \ 
{ \rm is \ independent \ of \ the \ (F \ or \  P)} \ BC_y \ . 
\label{alpha_Lambda}
\eeq
With this inference, we denote the resultant common limit for either of these
transverse boundary conditions as $\alpha(\Lambda)$, where $\Lambda$ refers to
the infinite lattice $\Lambda$.  

Because there is no transverse boundary to the
strip if one uses periodic transverse (cylindrical or toroidal) boundary
conditions, one expects that these boundary conditions yield values of
$\alpha(\Lambda,(L_y)_P \times \infty)$ that approach the $L_y=\infty$ value,
$\alpha(\Lambda)$, more rapidly than if one uses free transverse boundary
conditions and calculates the resultant $\alpha(\Lambda,(L_y)_F \times \infty)$
values.  Our results in \cite{ka3} and here are in agreement with this
expectation. Provided that this monotonicity holds for all higher values of
strip width $L_y$, it follows that the maximal value that we obtain for
$\alpha(\Lambda,(L_y)_P \times \infty)$ with the largest $L_y$ for which we
have performed the calculation is a lower bound for $\alpha(\Lambda)$. 
As we will discuss below, a comparison of our values of 
$\alpha(tri,(L_y)_P \times \infty)$ with the precise value of $\alpha(tri)$
that we calculate (see Eq. (\ref{alpha_tri})) gives further strong support to 
this inference. In order to measure the convergence of consecutive values of 
$\alpha(\Lambda,(L_y)_{BC_y} \times \infty)$ to a constant limiting value, we
define the ratio
\beq
R_{\alpha,\Lambda,(L_y+1)/L_y,BC_y} \equiv 
\frac{\alpha(\Lambda,(L_y+1)_{BC_y} \times \infty)}
     {\alpha(\Lambda,(L_y)_{BC_y} \times \infty)} \ . 
\label{ralpha}
\eeq
As will be evident from our results for the square, triangular, and honeycomb
lattices, even for modest values of the strip widths, these ratios approach
very close to unity.


\subsection{Strips of the Square Lattice}

In Table \ref{lowerbounds_alpha_sq_table} we list the values of
$\alpha(sq,(L_y)_F \times \infty)$ and $\alpha(sq,(L_y)_P \times \infty)$ that
we have calculated. The values of $\alpha(sq,(L_y)_F \times \infty)$ for $1 \le
L_y \le 8$ and of $\alpha(sq,(L_y)_P \times \infty)$ for $1 \le L_y \le 12$
were given in \cite{ka3}, while the value of $\alpha(sq,13_P \times \infty)$ is
new here. To show the convergence quantitatively to high accuracy, we have
listed the values of $\alpha(sq,(L_y)_F \times \infty)$ and $\alpha(sq,(L_y)_P
\times \infty)$ to more significant figures than were given in \cite{ka3}, and
we have also listed values of the ratio $R_{\alpha,sq,(L_y+1)/L_y,BC_y}$.
Using (\ref{alpha_monotonic}) and (\ref{alpha_Lambda}), we therefore infer the
lower bound
\beq
\alpha(sq) > 3.4932448 \ . 
\label{alpha_sq_lowerbound}
\eeq

This new lower bound may be compared with previous lower bounds on
$\alpha(sq)$. (In this context, it should be mentioned that our notation is
different from the notation used in Refs.
\cite{merino_welsh99}-\cite{garijo2014}; our quantities $a(G)$, $a_0(G)$, and
$b(G)$ are the same as their $\alpha(G)$, $\alpha_0(G)$, and $\beta(G)$,
respectively, and our quantities $\alpha(\{G\})$, $\alpha_0(\{G\})$, and
$\beta(\{G\})$ are the same as their quantities $\lim_{n(G) \to \infty}
\alpha(G)^{1/n(G)}$, $\lim_{n(G) \to \infty} \alpha_0(G)^{1/n(G)}$, and
$\lim_{n(G) \to \infty} \beta(G)^{1/n(G)}$, respectively.  In
\cite{merino_welsh99}, Merino and Welsh proved that
\beq
\frac{22}{7} \le \alpha(sq) \le 3.709259278  
\label{merino_welsh99_alpha}
\eeq
and
\beq
\frac{7}{3} \le \alpha_0(sq) \le 3.21 \ . 
\label{merino_welsh99_alpha0}
\eeq
In \cite{cmnn}, Calkin, Merino, Noble, and Noy proved more 
restrictive lower and upper bounds on $\alpha(sq)$, namely 
\beq
3.41358 \le \alpha(sq) \le 3.55449 \ . 
\eeq
Subsequently, in \cite{garijo2014}, Garijo et al. obtained still more
restrictive lower and upper bounds on $\alpha(sq)$, namely
\beq
3.42351 \le \alpha (sq) \le 3.5477 \ . 
\eeq
Evidently, the new lower bound (\ref{alpha_sq_lowerbound}) that we have
inferred is consistent with, and more restrictive than these previous lower
bounds. We will also infer a more restrictive upper bound on $\alpha_0(sq)$
below.


\section{Method to Infer Lower and Upper Bounds on Exponential Growth
  Constants}
\label{method_section}

In this section we explain a method that we use to infer lower and upper bounds
on the exponential growth constants $\alpha(\Lambda)$, $\alpha_0(\Lambda)$, and
$\beta(\Lambda)$ for several lattices $\Lambda$. Let us discuss
$\alpha(\Lambda)$ and $\alpha_0(\Lambda)$ first. We recall Eqs.
(\ref{alpha_infstrip}) and (\ref{alpha0_infstrip}). For definiteness, we
specialize the following discussion to strip graphs of the square
lattice. Corresponding results for other lattices are similar with appropriate
modifications. 

As discussed above, our results are consistent with the inference
(\ref{alpha_Lambda}) so that we may equivalently use strips with free or
periodic transverse boundary conditions (as well as free or periodic
longitudinal boundary conditions).  For strips of the square lattice with
width $L_y$, the values of $\alpha(sq)$ and $\alpha_0(sq)$ are thus given,
respectively, by the following, where, as before, $BC_y$ denotes the transverse
boundary condition, free or periodic (i.e., cylindrical)
\beqs
\alpha(sq) &=& \lim_{n \to \infty} [P([sq,(L_y)_{BC_y} \times m],-1)]^{1/n} = 
\lim_{L_y \to \infty} [\lambda_{sq,L_y,BC_y}(-1)]^{1/L_y} \cr\cr
 &=& \lim_{L_y \to \infty} \alpha(sq,(L_y)_{BC_y} \times \infty)
\label{alpha_sq}
\eeqs
and 
\beqs
\alpha_0(sq)&=&\lim_{n \to \infty} [P_r([sq,(L_y)_{BC_y} \times m],0)]^{1/n} = 
\lim_{L_y \to \infty} [\lambda_{sq,L_y,BC_y}(0)]^{1/L_y} \cr\cr
&=& \lim_{L_y \to \infty} \alpha_0(sq,(L_y)_{BC_y} \times \infty) \ .
\label{alpha0_sq}
\eeqs
Since $a(G)$ and $a_0(G)$ can be calculated from the chromatic polynomial
$P(G,q)$ without the necessity of calculating the full two-variable Tutte
polynomial or equivalent Potts model partition function, the $\lambda$
functions that will be used for our analysis are those that occur in $P(G,q)$
and hence depend on the single variable $q$. These are evaluated at $q=-1$ for
$a(G)$ and at $q=0$ for $a_0(G)$, and we indicate this in the notation. The
calculation of $b(G)$ requires an evaluation of the full Tutte polynomial,
$T(G,x,y)$ with $(x,y)=(0,2)$, or equivalently, the Potts model partition
$Z(G,q,v)$ with $(q,v)=(-1,1)$, and consequently in our discussion below of
$b(G)$ and the corresponding exponential growth constant, $\beta(\{G\})$, the
$\lambda$ functions involved will be those for the Tutte polynomial and hence
will depend on two variables. As noted above, for all of these exponential
growth constants, we do not actually need the full chromatic or Tutte
polynomial, but only the dominant $\lambda$ function.

From our explicit calculations for the full range of $L_y$ values that we have
investigated, we have observed that all the quantities $[\lambda_{sq, L_y,
  free}(-1)]^{1/L_y}$, $[\lambda_{sq, L_y, free}(0)]^{1/L_y}$, $[\lambda_{sq,
  L_y, cyl}(-1)]^{1/L_y}$, and $[\lambda_{sq, L_y,cyl}(0)]^{1/L_y}$ increase
monotonically as $L_y$ increases.  Provided that this monotonic increase
continues for larger $L_y$, our values thus yield lower bounds on the
respective asymptotic values in the limit as $L_y \to \infty$, i.e., the values
of $\alpha(\Lambda)$ and $\alpha_0(\Lambda)$ for the infinite lattices
$\Lambda$.  In \cite{ka3} we had used $[\lambda_{sq,L_y,cyl}(-1)]^{1/L_y}$ with
$L_y$ up to 12 to obtain the lower bound on $\alpha(sq)$ that we gave in that
paper.  

As an explicit example, we consider strips of the square lattice. For $L_y=2$,
the dominant $\lambda$ is
\beq
\lambda_{sq,2,free} = q^2-3q+3 \ . 
\label{lamsqly2}
\eeq
Thus, $\alpha(\{ L \}) = (\lambda_{sq,2,free})^{1/{L_y}} = 
\sqrt{\lambda_{sq,2,free}}$ evaluated at $q=-1$, which yields the result in 
Eq. (\ref{alpha_sqlad}) above. The corresponding evaluation at $q=0$ yields
the value in (\ref{alpha0_sqlad}).  

For $L_y=3$, depending on the value of $q$, the dominant $\lambda$ functions
are \cite{strip,wcyl,wcy}
\beq
\lambda_{sq,3,free,\pm} = \frac{1}{2} \Big [ (q-2)(q^2-3q+5) \pm 
\sqrt{(q^2-5q+7)(q^4-5q^3+11q^2-12q+8)} \, \Big ] \ . 
\label{lamsqly3}
\eeq
If (real) $q \ge 2$, then the function $\lambda_{sq,3,free,+}$ is dominant
(i.e., has the larger magnitude) and determines the $W$ function defined in
Eq. (\ref{w}) as $W = (\lambda_{sq,3,free,+})^{1/3}$. In contrast, for the
values of $q$ that are relevant here, namely $q=-1$ and $q=0$, the function
$\lambda_{sq,3,free,-}$ has the larger magnitude and hence is dominant. This
$\lambda_{sq,3,free,-}$ function is negative for $q < 2.685$, but, as is
evident in Eqs. (\ref{alpha_wqm1}) and (\ref{alpha0_wq0}), only the magnitude
is relevant for the $\alpha$ and $\alpha_0$ exponential growth constants. (For
finite-length strips, the factor $(-1)^n(G)$ in Eq.  (\ref{a_pqm1}) and the
factor $(-1)^{n(G)-1}$ in Eq. (\ref{a0_prq0}) yield positive values for $a(G)$
and $a_0(G)$.)  To avoid magnitude signs cluttering various equations, it is
understood implicitly that, where necessary, we remove minus signs so that the
dominant $\lambda$ function is positive for $q=-1$ and $q=0$.  In the present
case of the $L_y=3$ square-lattice strips, we thus set $|\lambda_{sq,3,free,-}|
\equiv \lambda_{sq,3,free}$, and similarly for other strips. Evaluating this at
$q=-1$ and $q=0$ and taking the $1/L_y=1/3$ root, we get
\beq
\alpha(sq,L_y=3,free) = \Bigg ( \frac{27 + \sqrt{481}}{2} \Bigg )^{1/3} = 
2.903043 
\label{alphasq_ly3}
\eeq
and
\beq
\alpha_0(sq,L_y=3,free) = (5 + \sqrt{14} \, )^{1/3} = 
2.0599875 \ . 
\label{alpha0sq_ly3}
\eeq
We observe the inequalities $\alpha(sq,3,free) >
\alpha(sq,2,free)$, where $\alpha(sq,2,free) \equiv \alpha(\{ L \})$ in
Eq. (\ref{alpha_sqlad}), and $\alpha_0(sq,3,free) >
\alpha_0(sq,2,free)$, where $\alpha_0(sq,2,free) \equiv \alpha_0(\{ L \})$ in
Eq. (\ref{alpha0_sqlad}). These are in accord with the monotonicity relations
that were noted above. 

The property that we find in our calculations, that $\alpha(\Lambda,L_y,free)$
and $\alpha_0(\Lambda,L_y,free)$ are monotonically increasing functions of
strip width for a strip of the lattice $\Lambda$, is opposite to the behavior
that was found for $W(\Lambda,L_y,free,q)$ for values of $q$ used in proper
$q$-colorings of the lattice $\Lambda$ \cite{bcc99}. This reversal can be
traced to the evaluation at different values of $q$, namely $q=-1$ and $q=0$
here, as contrasted with values of $q$ used for proper $q$ colorings of
$\Lambda$.

Related to this, we have noticed an interesting connection between our results
for $\alpha(\Lambda)$ and $\alpha_0(\Lambda)$ and the analytic expressions that
were proved in \cite{wn} (see also \cite{ww,w3}) to be lower bounds on
$W(\Lambda,q)$ for all Archimedean lattices (and dual Archimedean lattices)
using a coloring-matrix method introduced in \cite{biggscoloring} to prove a
lower bound on $W(sq,q)$.  We find that if one evaluates these expressions at
$q=-1$ and $q=0$, then the results are consistent with being upper bounds on
$\alpha(\Lambda)$ and $\alpha_0(\Lambda)$, respectively.  As discussed in the
Appendix, an Archimedean lattice $\Lambda$ has the form $\Lambda = (\prod_i
p_i^{a_i})$, where this product refers to the ordered sequence of polygons
traversed in a circuit around any vertex, and the $i$'th polygon has $p_i$
sides, appearing $a_i$ times contiguously in the sequence (it can also occur
non-contiguously). The total number of occurrences of the polygon $p_i$ in the
above sequence is denoted as $a_{i,s}$. In this general notation, $(sq)=(4^4)$,
$(tri)=(3^6)$, and $(hc)=(6^3)$. The number of polygons of type $p_i$ per
vertex is
\beq
\nu_{p_i} = \frac{a_{i,s}}{p_i} \ . 
\label{nupi}
\eeq
Note that $\nu_{p_i}$ coincides with $\nu_{\Lambda}$ in Eq. (\ref{nu_g}) 
and takes on the values in Eqs. (\ref{nusq})-(\ref{nuhctri}) for the square, 
triangular, and honeycomb lattices. The lower bound proved in
\cite{wn} for a general Archimedean lattice is
\beq
W(\Lambda,q) \ge W(\Lambda,q)_\ell \ , 
\label{wwlow}
\eeq
where 
\beq
W \bigg ( (\prod_i p_i^{a_i} ),q \bigg )_\ell = 
\frac{\prod_i [D_{p_i}(q)]^{\nu_{p_i}}}{q-1} \ , 
\label{wlowform}
\eeq
with
\beq
D_n(q) = \sum_{s=0}^{n-2} (-1)^s {n-1 \choose s} q^{n-2-s} \ .
\label{dn}
\eeq
This lower bound applies for $q \ge \chi(G)$, where $\chi(G)$ is the chromatic
number of $G$ (i.e., the minimum value of $q$ required for a proper
$q$-coloring of $G$).  In \cite{wn,ww,w3} this lower bound on $W(\Lambda,q)$ was shown
to be very close to the actual values of $W(\Lambda,q)$ as determined from
Monte-Carlo measurements, large-$q$ series, and the exact result for $W(tri,q)$
from \cite{baxter86,baxter87}. We conjecture that for each Archimedean lattice
$\Lambda$, the evaluation of the right-hand side of Eq. (\ref{wlowform}) at the
respective values $q=-1$ and $q=0$ yields respective upper bounds on
$\alpha(\Lambda)$ and $\alpha_0(\Lambda)$. which would read
\beq
\alpha(\Lambda) < \alpha_{u,w}(\Lambda)
\label{alpha_lt_alphaw}
\eeq
and 
\beq
\alpha_0(\Lambda) < \alpha_{0,u,w}(\Lambda) \ , 
\label{alpha0_lt_alpha0w}
\eeq
where 
\beq
\alpha_{u,w}(\Lambda) = \frac{\prod_i |D_{p_i}(-1)|^{\nu_{p_i}}}{2}
\label{alfup_uw}
\eeq
and
\beq
\alpha_{0,u,w}(\Lambda) = \prod_i |D_{p_i}(0)|^{\nu_{p_i}} \ . 
\label{alf0up_uw}
\eeq
To indicate the connection with the $W(\Lambda,q)$ bounds, we use a subscript
$w$ in these expressions. Specifically, for the three
Archimedean lattices under consideration here, in order of increasing vertex
degree $\Lambda(\Lambda)$, these conjectured upper bounds are 
\beq
\alpha_{u,w}(hc) = \frac{\sqrt{31}}{2} = 2.78388218
\label{alpha_hc_upper_wn}
\eeq
\beq
\alpha_{u,w}(sq) = \frac{7}{2} 
\label{alpha_sq_upper_wn}
\eeq
\beq
\alpha_{u,w}(tri) = \frac{9}{2} 
\label{alpha_tri_upper_wn}
\eeq
and
\beq
\alpha_{0,u,w}(hc) = \sqrt{5} = 2.236068
\label{alpha0_hc_upper_wn}
\eeq
\beq
\alpha_{0,u,w}(sq) = 3
\label{alpha0_sq_upper_wn}
\eeq
and
\beq
\alpha_{0,u,w}(tri) = 4 \ . 
\label{alpha0_tri_upper_wn}
\eeq
As will be seen below, these are close to the upper bounds that we derive from
our studies of strip graphs of these lattices, and to the exact values that we
obtain for $\alpha(tri)$ and $\alpha_0(tri)$. (The respective values in
Eqs. (\ref{alpha_hc_upper_wn})-(\ref{alpha0_tri_upper_wn}) are the 
$(L_y+1)/L_y=2/1$ entries in the corresponding tables based on strip
graph calculations.)  A plausible inference is that this closeness of 
the values (\ref{alpha_hc_upper_wn})-(\ref{alpha0_tri_upper_wn}) to the optimal
inferred upper bounds is related to the fact that the lower bound 
(\ref{wwlow})-(\ref{dn}) that was rigorously proved in \cite{wn} for all 
Archimedean lattices is very close to the actual values of $W(\Lambda,q)$ on 
these lattices.

From \cite{a}, the dominant $\lambda$ function in $Z(G,q,v)$ for the relevant 
values $q=-1$ and $v=1$ is 
\beq
\lambda_{sq,2,free}(q,v) = \frac{v}{2} \Big [ q + v(v+4) + 
\sqrt{ v^4 + 4v^3+12v^2 -2qv^2 +4qv+q^2} \, \Big ]
\label{zlamsq}
\eeq
with the evaluation 
\beq
\lambda_{sq,2,free} = 4 \quad {\rm at} \ (q,v) = (-1,1) \ . 
\label{zlamsqb}
\eeq
Taking the $1/(L_y)=1/2$ root, one obtains the result for 
$\beta(sq,2,free) \equiv \beta(\{ L \})$ given above in
Eq. (\ref{beta_sqlad}).

From \cite{s3a} we find that for the $L_y=3$ strip of the square lattice with
free transverse boundary conditions, the dominant $\lambda$ at $(q,v)=(-1,1)$
is a root of the sixth-degree equation given in Eqs. (A.1)-(A.7) of
\cite{s3a}, with the value
\beq
\lambda_{sq,3,free}=\frac{17+\sqrt{145}}{2} \quad {\rm at} \ (q,v) = (-1,1) 
\ .
\label{zlamly3b}
\eeq
Taking the 1/3 root, one obtains the result 
\beq
\beta(sq,3,free)= \Bigg ( \frac{17+\sqrt{145}}{2} \, \Bigg )^{1/3} 
= 2.439665 \ .
\label{beta_ly3sq}
\eeq
We note the inequality $\beta(sq,3,free) > \beta(sq,2,free)$, in agreement
with the general monotonicity property noted above.

We next prove a useful inequality. For this purpose, we begin by 
considering lattice strip graphs with width $L_y=2^p$ for some (positive) 
integer power $p$.  This inequality applies to the
exponential growth constant $\phi$ for the Tutte polynomial of a
recursive family of graphs (e.g., lattice strip graphs) for 
$x \ge 0$ and $y \ge 0$, where $\phi$ is defined as 
\beq
\phi( \{ G \},x,y) = \lim_{n(G) \to \infty} [T(G,x,y)]^{1/n(G)} \ . 
\label{taudef}
\eeq
If an edge $e \in E$ is not a loop or a
bridge, then the Tutte polynomial satisfies the deletion-contraction relation
\beq
T(G,x,y) = T(G-e,x,y) + T(G/e,x,y) \ , 
\label{dcr}
\eeq
where $G-e$ denotes $G$ with the edge $e$ deleted and $G/e$ denotes the result
of deleting the edge $e$ from $G$ and identifying the vertices that this edge
connected.  (For a graph that is comprised of $\ell$ loops and $b$ bridges,
$T(G,x,y)=x^by^\ell$. The proof of the inequality follows from an iterative
use of the deletion-contraction relation. This leads to nested inequalities for
the dominant $\lambda$ function for all of the three cases $(x,y)=(2,0)$ for
$\alpha(\{G\})$, $(x,y)=(1,0)$ for $\alpha_0(\{G \})$, and $(x,y)=(0,2)$ for
$\beta(\{G \})$.  The basic observation is that if one compares the Tutte
polynomial for, say, an $L_x \times 4$ strip of a lattice $\Lambda$, with the
Tutte polynomial for a (disconnected) graph consisting of two copies of an $L_x
\times 2$ strip, then the former has $L_x$ more edges, the removal of which
yields the the latter two graphs. By iterative application of the
deletion-contraction theorem, one can relate the free strip of width $L_y=4$ to
the graph consisting of two $L_y=2$ free strips, and the inequality then
follows. 

From Eqs. (\ref{alpha_infstrip}) and (\ref{alpha0_infstrip}), it follows that
for the square and triangular lattices, 
as the strip width $L_y \to \infty$, $\lambda(\Lambda,(L_y)_{BC_y} \times
\infty)(-1) \sim [\alpha(\Lambda)]^{L_y}$ and
$\lambda(\Lambda,(L_y)_{BC_y} \times \infty)(0) \sim 
[\alpha_0(\Lambda)]^{L_y}$.  Therefore, another measure of the asymptotic
large-$L_y$ limit is given by the ratio
\beq
\frac{\lambda(\Lambda,(L_y)_{BC_y} \times \infty)(q)}
     {\lambda(\Lambda,(L_y-1)_{BC_y} \times \infty)(q)} \ , 
\label{lamrat1}
\eeq
where $q=-1$ for $\alpha(\Lambda)$ and where $q=0$ for $\alpha_0(\Lambda)$. 
Illustrating this with our illustrative strip, we observe that the ratios 
of the dominant $\lambda(q)$ functions in the chromatic polynomial at $q=-1$
and $q=0$ are 
\beq
\frac{\lambda(sq,3_F \times \infty)(-1)}
     {\lambda(sq,2_F \times \infty)(-1)} = \frac{27+\sqrt{481}}{14}= 3.495122
\label{lamsq32_alpha}
\eeq
and
\beq
\frac{\lambda(sq,3_F \times \infty)(0)}
     {\lambda(sq,2_F \times \infty)(0)} = \frac{5+\sqrt{14}}{3} = 2.913886 \ . 
\label{lamsq32_alpha0}
\eeq
(The equivalent notation
$\frac{\lambda(sq,3,free)(-1)}{\lambda(sq,2,free)(-1)}$ and
$\frac{\lambda(sq,3,free)(0)}{\lambda(sq,2,free)(0)}$ is used in the tables.)

A corresponding discussion applies for $\beta(\Lambda)$, with the dominant
$\lambda(q)$ function in the chromatic polynomial replaced with the dominant
$\lambda(q,v)$ function in the Potts/Tutte polynomial, evaluated at
$(x,y)=(0,2)$ i.e., $(q,v)=(-1,1)$. For the honeycomb lattice, one replaces
these ratios by square roots. Thus, for the dominant $\lambda(q,v)$ functions
in $Z(G,q,v)$ at $(q,v)=(-1,1)$ on the $L_y=2$ and $L_y=3$ strips of the square
lattice with free transverse boundary conditions, we have
\beq
\frac{\lambda(sq,3_F \times \infty)(-1,1)}
     {\lambda(sq,2_F \times \infty)(-1,1)} = 
\frac{17+\sqrt{145}}{8} = 3.630199 \ . 
\label{lamsq32_beta}
\eeq
(The equivalent notation 
$\frac{\lambda(sq,3,free)(-1,1)}{\lambda(sq,2,free)(-1,1)}$ is used in the 
tables.)

Now consider the special case of the exponential growth constant $\alpha(\{G
\})$, for which we actually only need the one-variable special case of the
Tutte polynomial given by the chromatic polynomial, with the associated
$\lambda$ functions evaluated at $q=-1$ for the discussion of acyclic
orientations, as will be indicated in the notation below. We then have the
sequence of inequalities
\beqs
\lambda_{sq,1,free}(-1) &&\le [\lambda_{sq,2,free}(-1)]^{1/2} \le
[\lambda_{sq,4,free}(-1)]^{1/4} \le [\lambda_{sq, 8, free}(-1)]^{1/8} 
\cr\cr
&& \le ... \le \lim_{L_y \to \infty} [\lambda_{sq, L_y, free}(-1)]^{1/L_y} \ . 
\eeqs
Let us focus on one of these inequalities, namely $[\lambda_{sq, 2,
  free}(-1)]^{1/2} \le [\lambda_{sq, 4, free}(-1)]^{1/4}$.  The others can be
treated in a similar manner. Since $[\lambda_{sq,2,free}(-1)]^{L_x}$ is the
dominant $\lambda$ function for the chromatic polynomial of the $2 \times L_x$
strip, and equivalently of the Tutte polynomial with $(x,y)=(2,0)$, it
determines the corresponding $\phi$ function in the limit $L_x \to \infty$,
while $[\lambda_{sq,4,free}(-1)]^{L_x}$ similarly gives the $\phi$ function for
the $L_x \to \infty$ limit of the Tutte polynomial of the $4 \times L_x$ strip.
Now compare two $L_y=2$ strips with a $L_y=4$ strip. The former has $L_x$ fewer
edges than the latter, so the Tutte polynomial of the former is smaller than
that of the latter, since the coefficients of the Tutte polynomial (in terms of
variables $x$ and $y$) are positive. That is, 
$[\lambda_{sq,2,free}(-1)]^{2L_x} \le [\lambda_{sq,4,free}(-1)]^{L_x}$. 
This completes the proof. By the same type of argument, it follows, for
example, that
\beqs
\lambda_{sq, 1, free}(-1) &&\le [\lambda_{sq, 3, free}(-1)]^{1/3} \le
[\lambda_{sq, 6,free}(-1)]^{1/6} \le [\lambda_{sq, 12, free}(-1)]^{1/12} 
\cr\cr
&&\le ... \le \lim_{L_y \to \infty} [\lambda_{sq, L_y, free}(-1)]^{1/L_y} \ ,
\eeqs
where here $L_y=3 \cdot 2^s$, where $s$ is a non-negative integer. Other
corresponding inequalities with larger values of $L_y$ follow in the same way.

It is easy to prove that
\beq
\lambda_{sq, L_y, free}(-1) \le \lambda_{sq, L_y, cyl}(-1) \ , 
\label{ineq10}
\eeq
as follows. Consider an assignment of arrows on all of the edges of a free
strip of the square lattice with width $L_y$, such that there are no cycles,
i.e., an acyclic orientation of this strip. We can add $L_x$ more edges to
produce the corresponding cylindrical strip of the square lattice with the same
width $L_y$. Now we assign an orientation for the arrow on each of these
directed edges. If a choice of the direction of the arrow would result in a
cycle, then we choose the opposite direction for the arrow. It is impossible
that both choices will result a cycle, since that would mean that there would
already have been a cycle in the original free strip, which would contradict
the beginning assumption of an acyclic orientation.  This statement applies to
each of the additional $L_x$ edges of the cylindrical strips, so the number of
acyclic orientations on the cylindrical strip is at least the same as the
number on the free strip.  Therefore, for a strip graph of the lattice
$\Lambda$, the quantity $[\lambda_{sq, L_y, cyl}(-1)]^{1/L_y}$ serves as a
better lower bound on $\alpha(\Lambda)$ than $[\lambda_{sq,
  L_y,free}(-1)]^{1/L_y}$.  Alternatively, one can again use the
deletion-contraction relation (\ref{dcr}) to prove (\ref{ineq10}), since the
strip with cylindrical boundary conditions has $L_x$ more edge than the strip
with free boundary conditions with the same $L_y$.  Similar discussions apply
for acyclic orientations with a unique source, and for the $tri$ and $sq_d$
lattices.  For the honeycomb lattice, it is $[\lambda_{hc,
  L_y,free}(-1)]^{1/(2L_y)}$ that yields the corresponding values of $\alpha$
and $\alpha_0$ with $q=-1$ and $q=0$, respectively.  We thus infer the two
inequalities
\beq
\alpha(\Lambda) > \alpha(\Lambda,(L_y)_P \times \infty) \quad {\rm for \
  the \ maximal \ calculated \ value \ of} \ L_y 
\label{alpha_lowerbound}
\eeq
and
\beq
\alpha_0(\Lambda) > \alpha_0(\Lambda,(L_y)_P \times \infty) \quad {\rm for \
  the \ maximal \ calculated \ value \ of} \ L_y \ ,
\label{alpha0_lowerbound}
\eeq
where the right-hand sides of these inequalities are given, respectively, by
$[\lambda_{\Lambda, L_y, cyl}(-1)]^{1/L_y}$ and $[\lambda_{\Lambda, L_y,
  cyl}(0)]^{1/L_y}$ for $\Lambda=sq, \ tri, \ sq_d$, and by the corresponding
square roots of these functions for $\Lambda = hc$.  The corresponding bounds
also apply for free transverse boundary conditions, but, as noted, the bounds
with periodic transverse boundary conditions are more restrictive.

As mentioned above, we have shown by explicit calculation that
\beq
\lambda_{sq,1,free}(-1) < [\lambda_{sq,2,free}(-1)]^{1/2} < 
[\lambda_{sq,3,free}(-1)]^{1/3} < ... <  [\lambda_{sq,8,free}(-1)]^{1/8}
\label{sqlower}
\eeq
for the square lattice. This sequence should approach $\alpha(sq)$ as the strip
width $L_y \to \infty$. With the inference that 
\beq
[\lambda_{sq, L_y, free}(-1)]^{1/L_y} < 
[\lambda_{sq,L_y+1,free}(-1)]^{1/(L_y+1)} \ , 
\eeq
this is equivalent to
\beq
\lambda_{sq, L_y, free}(-1) < [\lambda_{sq, L_y+1, free}(-1)]^{L_y/(L_y+1)}
\eeq
and
\beq
[\lambda_{sq, L_y+1, free}(-1)]^{1/(L_y+1)} < 
\frac{\lambda_{sq, L_y+1, free}(-1)}{\lambda_{sq, L_y, free}(-1)} \ .
\eeq
From our explicit calculation, we find that
\beq
\frac{\lambda_{sq, 8, free}(-1)}{\lambda_{sq, 7, free}(-1)} <
\frac{\lambda_{sq, 7, free}(-1)}{\lambda_{sq, 6, free}(-1)} 
< ... < \frac{\lambda_{sq, 3,free}(-1)}{\lambda_{sq, 2, free}(-1)} 
< \frac{\lambda_{sq, 2, free}(-1)}{\lambda_{sq, 1, free}(-1)} \ .
\eeq
This leads us to infer that the ratio 
$\lambda_{sq, L_y+1, free}(-1)/\lambda_{sq, L_y, free}(-1)$ serves as an 
upper bound for $\alpha(sq)$.  From Eq. (\ref{alpha_infstrip}) and
the proof above that $[\lambda_{sq, L_y, free}(-1)]^{1/L_y}$ approaches 
$\alpha(sq)$ from below (and is very close to it when $L_y >> 1$), one 
could infer that $\lambda_{sq, L_y+1, free}(-1)$ is close to 
$[\alpha(sq)]^{L_y+1}$ and $\lambda_{sq, L_y, free}(-1)$ is also close 
(but not as close) to $[\alpha(sq)]^{L_y}$. Therefore, 
$\frac{\lambda_{sq, L_y+1, free}(-1)}{\lambda_{sq, L_y, free}(-1)}$ 
should be slightly larger than $\alpha(sq)$, and hence should serve as an 
upper bound on $\alpha(sq)$. Similar discussions apply for
the evaluation at $q=0$ and for the $tri$ and $sq_d$ lattices.  For the
honeycomb lattice, one replaces this ratio by its square root to obtain the
upper bound on $\alpha(hc)$. We thus infer the two inequalities
\beq
\alpha(\Lambda) < \frac{\lambda_{\Lambda, L_y+1, free}(-1)}
{\lambda_{\Lambda, L_y, free}(-1)}  \quad {\rm for \
  the \ maximal \ calculated \ value \ of} \ L_y 
\label{alpha_upperbound}
\eeq
and
\beq
\alpha_0(\Lambda) < \frac{\lambda_{\Lambda, L_y+1, free}(0)}
{\lambda_{\Lambda, L_y, free}(0)} \quad {\rm for \
  the \ maximal \ calculated \ value \ of} \ L_y \ ,
\label{alpha0_upperbound}
\eeq
for $\Lambda=sq, \ tri, \ sq_d$ lattices.  We infer the corresponding 
inequalities for the honeycomb lattice with the ratios on the right-hand sides
replaced by their square roots. 

Another argument that supports this inference is the following, where we again
specialize to the square-lattice strips for definiteness. 
Let us define the ratio of the adjacent terms in (\ref{sqlower}) (i.e.,
successive lower bounds on $\alpha(sq)$ from the infinite-length strip of width
$L_y$ and width $L_y-1$) as
\beq
R_{sq, \frac{L_y}{L_y-1}, free}(-1) \equiv 
\frac{[\lambda_{sq, L_y, free}(-1)]^{1/L_y}}
{[\lambda_{sq, L_y-1, free}(-1)]^{1/(L_y-1)}} \ .
\eeq
This ratio $R_{sq, (L_y+1)/L_y, free}(-1)$ is larger than 1. We find that this
ratio decreases toward 1 from above as $L_y$ increases from 1 to 7, as listed
in the next section. The same statement applies when the boundary condition is
cylindrical. This is consistent with the inference that our lower bound is
approaching an asymptotic constant value, namely the value for the infinite
lattice.

Provided that this property continues to hold for any $L_y$, namely, 
\beq
\frac{[\lambda_{sq, L_y+1, free}(-1)]^{1/(L_y+1)}}
     {[\lambda_{sq, L_y,   free}(-1)]^{1/L_y}} < 
\frac{[\lambda_{sq, L_y, free}(-1)]^{1/L_y}}
     {[\lambda_{sq, L_y-1,free}(-1)]^{1/(L_y-1)}} \ ,
\eeq
then it is equivalent to
\beqs
[\lambda_{sq,L_y+1,free}(-1)]^{1/(L_y+1)} & < & 
\frac{[\lambda_{sq, L_y,free}(-1)]^{2/L_y}}
     {[\lambda_{sq,L_y-1,free}(-1)]^{1/(L_y-1)}} 
\cr
& = & \frac{\lambda_{sq, L_y, free}(-1)}{\lambda_{sq, L_y-1, free}(-1)}
\times 
\frac{[\lambda_{sq, L_y-1, free}(-1)]^{(L_y-2)/(L_y-1)}}
{[\lambda_{sq, L_y, free}(-1)]^{(L_y-2)/L_y}} \cr
& < & \frac{\lambda_{sq, L_y, free}(-1)}{\lambda_{sq, L_y-1, free}(-1)} \ .
\eeqs
Let us define the ratio 
\beq
R_{sq, \frac{L_y^2}{(L_y-1)(L_y+1)}, free}(-1) \equiv 
\frac{[\lambda_{sq, L_y, free}(-1)]^2} 
{\lambda_{sq, L_y-1, free}(-1) \lambda_{sq, L_y+1, free}(-1)} \ .
\eeq
This is the ratio of adjacent upper bounds. Since the upper bounds decrease as
the strip width $L_y$ increases, the larger-$L_y$ upper bound in the
denominator is smaller than the smaller-$L_y$ upper bound in the numerator, so
this ratio is also larger than unity.  We find that this ratio also 
decreases as $L_y$ increases from 2 to 7, as listed in the next section. This
is consistent with our upper bounds approaching a constant value as $L_y \to
\infty$, namely the value of $\alpha(sq)$ for the infinite lattice. 

Next, we consider the totally cyclic orientations on strip graphs and the
exponential growth constant $\beta(\Lambda)$ of the lattice $\Lambda$.  As
stated above in Eqs. (\ref{b_tx0y2}) and (\ref{bzqm1v1}), for a finite graph,
$G$, the number of totally cyclic orientations, $b(G)$, is given by the
evaluations $b(G)=T(G,0,2)$, or equivalently, by $b(G)=-Z(G,-1,1)$.  From our
earlier calculations of Potts model partition functions for strip graphs of
various lattices with cyclic and toroidal boundary conditions, we showed that
the dominant $\lambda$ function in the Potts partition function evaluated at
$(q,v)=(-1,1)$ has the coefficient $c^{(d)}=q-1$, in
the notation of Eq. (\ref{zcyc}).  Related to this, the result for $\beta$ in
the limit of infinite length depends on both the longitudinal and transverse
boundary conditions of the strip.  To minimize finite-size effects, we
therefore restrict to strips with periodic longitudinal boundary conditions in
our analysis of the $\beta$ exponential growth constant. 
The dominant $\lambda$ functions of this type will
be denoted as $\lambda_{sq, L_y, cyc}(q=-1,v=1) \equiv \lambda_{sq, L_y,
  cyc}(-1,1)$ and $\lambda_{sq, L_y,tor}(q=-1,v=1) \equiv \lambda_{sq,
  L_y,tor}(-1,1)$, respectively. We have obtained inequalities similar to those
discussed above in this case also.  That is, $[\lambda_{sq, L_y,
  cyc}(-1,1)]^{1/L_y}$ and $[\lambda_{sq, L_y, tor}(-1,1)]^{1/L_y}$ increase
monotonically as $L_y$ increases, and
\beq
\lambda_{sq, L_y, cyc} (-1,1) \le \lambda_{sq, L_y, tor} (-1,1) \ .
\eeq
We find similar results for other lattices.  This leads us to infer that 
$[\lambda_{\Lambda, L_y, tor}(-1,1)]^{1/L_y}$ is a lower
bound for $\beta(\Lambda)$ for $\Lambda=sq, \ tri, \ sq_d$:
\beq
\beta(\Lambda) > [\lambda_{\Lambda, L_y, tor}(-1,1)]^{1/L_y} \quad 
{\rm for \ the \ maximal \ calculated \ value \ of} \ L_y \ . 
\label{beta_lowerbound}
\eeq
for these lattices.  For the honeycomb lattice, we infer that the corresponding
lower bound holds with the right-hand side replaced by its square root, i.e.,
with the power $1/(2L_y)$ rather than $1/L_y$. We define the ratio
\beq
R_{sq, \frac{L_y}{L_y-1}, BC_y}(-1,1) \equiv 
\frac{[\lambda_{sq,L_y,BC_y}(-1,1)]^{1/L_y}}
     {[\lambda_{sq,L_y-1,BC_y}(-1,1)]^{1/(L_y-1)}}
\label{ratio10}
\eeq
for the adjacent lower bounds, where $BC_y$ can be either cyclic or toroidal
boundary conditions. For the honeycomb lattice, $L_y$ can only be an even
number for the strips with cylindrical or toroidal boundary conditions, and the
ratio analogous to (\ref{ratio10}) is defined by the results for strips with
width $L_y$ and $L_y-2$ rather than $L_y$ and $L_y-1$. .

We also find that the ratio 
\beq
R_{sq, \frac{L_y^2}{(L_y-1)(L_y+1)}, cyc} (-1,1) \equiv 
\frac{[\lambda_{sq, L_y, cyc}(-1,1)]^2} 
{\lambda_{sq, L_y-1, cyc} (-1,1) \lambda_{sq, L_y+1, cyc} (-1,1)} 
\eeq 
decreases when $L_y$ increases and therefore infer that 
$\lambda_{sq, L_y+1, cyc} (-1,1)/\lambda_{sq, L_y, cyc} (-1,1)$ provides 
an upper bound on $\beta(sq)$.  We find similar behavior for other lattices and
thus infer the upper bound 
\beq
\beta(\Lambda) < \frac{\lambda_{\Lambda, L_y+1, cyc} (-1,1)}
{\lambda_{\Lambda, L_y, cyc} (-1,1)} \quad 
{\rm for \ the \ maximal \ calculated \ value \ of} \ L_y \ . 
\label{beta_upperbound}
\eeq
for $\Lambda=sq, \ tri, \ sq_d$.  For the honeycomb lattice, $\Lambda=hc$, we
infer the corresponding inequality with the ratio on the right-hand side
replaced by its square root. 


\section{Numerical Values of Lower and Upper Bounds for $\alpha(\Lambda)$ 
and $\alpha_0(\Lambda)$} 
\label{alfalf0_values_section}

In this section we present our results for numerical values of lower and upper
bounds for $\alpha(\Lambda)$ and $\alpha_0(\Lambda)$ on various two-dimensional
lattices $\Lambda$.  For a given lattice $\Lambda$, we denote our lower
($\ell$) and upper ($u$) bounds with respective subscripts $\ell$ and $u$ as
$\alpha_\ell(\Lambda)$, $\alpha_u(\Lambda)$, $\alpha_{0,\ell}(\Lambda)$, and
$\alpha_{0,u}(\Lambda)$. 
Since we use the entries with the highest values of strip width $L_y$ for our
lower and upper bounds, we quote these to slightly higher precision than the
smaller-$L_y$ entries.  As noted above, we obtain our best lower and
upper bounds from the strips with periodic transverse boundary conditions. 
To begin, we present these results for the square lattice
in Tables \ref{lowerbounds_alpha_sq_table}-\ref{upperbounds_alpha0_sq_table}.

Next, for the triangular lattice, we can use Eqs. (\ref{alpha_wqm1}) and
(\ref{alpha0_wq0}) together with exact expressions for the $W$ function on the
triangular lattice from \cite{baxter86,baxter87} to obtain precise values of
$\alpha(tri)$ and $\alpha_0(tri)$. As discussed below, by duality, one thus
obtains a precise value of $\beta(hc)$.  These results provide a quantitative
measure of how close our lower and upper bounds are to the exact values and
show the very high degree of precision that we achieve with these bounds for
the square, triangular, and honeycomb lattices, even with modest values of
strip width $L_y$. For the relevant range of real $q \le 0$, with
$q=2-\xi-\xi^{-1}$, an infinite-product expression for $W(tri,q)$, applicable
for $0 \le \xi \le 1$, was given in \cite{baxter86,baxter87}, from which one
has
\beq
|W(tri,q)| = \frac{1}{\xi} \prod_{j=1}^\infty 
\frac{(1-\xi^{6j-3})(1-\xi^{6j-2})^2(1-\xi^{6j-1})}
{(1-\xi^{6j-5})(1-\xi^{6j-4})(1-\xi^{6j})(1-\xi^{6j+1})} \ . 
\label{wtri_qle0}
\eeq
We have evaluated this infinite product numerically for $q=-1$, i.e., 
$\xi = (3-\sqrt{5})/2$, using Maple and Mathematica. We obtain
\beq
\alpha (tri) = |W(tri,-1)| = 4.47464730907 \ . 
\label{alpha_tri}
\eeq
The values that we obtain from Maple and Mathematica serve to check each other
and agree with each other; their precision extends well beyond the twelve
significant figures listed in Eq. (\ref{alpha_tri}), but the numerical result
in Eq. (\ref{alpha_tri}) will be sufficient for our present purposes.

We combine an analytic evaluation of $W(tri,q)$ at $q=0$ from \cite{baxter86}
with the relation (\ref{alpha0_wq0}) to obtain an exact analytic result for
$\alpha_0(tri)$, namely 
\beq
\alpha_0 (tri) = |W(tri,0)| = 
\frac{3^{3/2} \, [\Gamma(\frac{1}{3})]^9}{(2\pi)^5}
= \frac{(2\pi)^4}{3^3 [\Gamma(\frac{2}{3})]^9} = 3.77091969752 \ , 
\label{alpha0_tri}
\eeq
where $\Gamma(z)$ is the Euler gamma function. The equality of the two analytic
expressions on the right-hand side of Eq. (\ref{alpha0_tri}) follows from the
reflection formula $\Gamma(z)\Gamma(1-z) = \pi/[\sin(\pi z)]$ with $z=1/3$. 

We list the ratios of lower and upper bounds to these exact values in Tables
\ref{lowerbounds_alpha_tri_table}-\ref{upperbounds_alpha0_tri_table}. 
For the triangular lattice, $\epsilon(tri)=8$, so we also obtain the exact
results
\beq
\rho_\alpha(tri) = 0.55933091363
\label{rho_alpha_tri}
\eeq
and
\beq
\rho_{\alpha_0}(tri) = \frac{2\pi^4}{3^3[\Gamma(\frac{2}{3})]^9} 
= 0.47136496219
\label{rho_alpha0_tri}
\eeq

We present our results on lower and upper bounds on $\alpha(hc)$ and
$\alpha_0(hc)$ in Tables
\ref{lowerbounds_alpha_hc_table}-\ref{upperbounds_alpha0_hc_table}. 

Summarizing the lower and upper bounds for the these lattices $\Lambda$, 
listed in order of increasing (uniform) vertex degree, $\Delta(\Lambda)$, we 
have, for $\alpha(\Lambda)$, the bounds, given to the indicated number of
significant figures (and, where available, the exact results): 
\beq
2.782197008 < \alpha(hc) < 
2.783486470 
\label{alpha_bounds_hc}
\eeq
\beq
3.493244874 < \alpha(sq) < 
3.493927960
\label{alpha_bounds_sq}
\eeq
and
\beq
4.471898355 < \alpha(tri) < 
4.474676977  
\label{alpha_bounds_tri}
\eeq
(with the precise value $\alpha(tri)=4.47464731$ in Eq. (\ref{alpha_tri}) 
from the exact result).  For $\alpha_0(\Lambda)$ we have
\beq
2.106218408 < \alpha_0(hc) < 
2.161128567 
\label{alpha0_bounds_hc}
\eeq
\beq
2.830007782 < \alpha_0(sq) < 
2.862213752
\label{alpha0_bounds_sq}
\eeq
and
\beq
3.737371971 < \alpha_0(tri) < 
3.780466270 
\label{alpha0_bounds_tri}
\eeq
(with the precise value $\alpha_0(tri)=3.7709196975$ from the exact result in
Eq. (\ref{alpha0_tri})).  Aside from $\alpha(tri)$ and $\alpha_0(tri)$, for
which we have given exact results, these lower and upper bounds are, to our
knowledge, the best current bounds on these exponential growth constants.  

The exact values of $\alpha(tri)$ and $\alpha_0(tri)$, and our upper bounds on
these other exponential growth constants, are close to the conjectured upper
bounds in Eq. (\ref{alpha_lt_alphaw})-(\ref{alpha0_tri_upper_wn}), especially
for $\alpha_u(\Lambda)$, as in evident from the following ratios (note that we
use the exact values of $\alpha(tri)$ and $\alpha_0(tri)$ in the numerators of
Eqs. (\ref{alpha_tri_over_w}) and (\ref{alpha0_tri_over_w}): 
\beq
\frac{\alpha_u(hc)}{\alpha_{u,w}(hc)} = 0.999858
\label{alpha_upperbound_hc_over_w}
\eeq
\beq
\frac{\alpha_u(sq)}{\alpha_{u,w}(sq)} = 0.998265
\label{alpha_upperbound_sq_over_w}
\eeq
\beq
\frac{\alpha(tri)}{\alpha_{u,w}(tri)} = 0.994366
\label{alpha_tri_over_w}
\eeq
\beq
\frac{\alpha_{0,u}(hc)}{\alpha_{0,u,w}(hc)} = 0.966486
\label{alpha0_upperbound_hc_over_w}
\eeq
\beq
\frac{\alpha_{0,u}(sq)}{\alpha_{0,u,w}(sq)} = 0.954071
\label{alpha0_upperbound_sq_over_w}
\eeq
and
\beq
\frac{\alpha_0(tri)}{\alpha_{0,u}(tri)} = 0.942730 \ . 
\label{alpha0_tri_over_w}
\eeq

A very important property of our lower and upper bounds on these exponential
growth constants is that they are quite close to each other. To show this 
quantitatively, we first calculate the average of
these values for each exponential growth constant (EGF),
\beq
EGC_{ave}(\Lambda) = \frac{(EGC)_\ell(\Lambda)+(EGF)_u(\Lambda)}{2}  \ , 
\label{egc_average}
\eeq
and then calculate the
fractional difference between the upper and lower bounds, i.e., the difference
divided by the average of these bounds,
\beq
\frac{EGC_u(\Lambda) - EGC_\ell(\Lambda)}{EGC_{ave}(\Lambda)} \ , 
\label{egc_fracdif}
\eeq
where EGC $=\alpha$, $\alpha_0$ (and $\beta$, as discussed below).

In order of increasing vertex degree, we obtain, for $\alpha(\Lambda)$,

\beq
\frac{\alpha_u(hc) - \alpha_\ell(hc)}{\alpha_{ave}(hc)} = 0.463 \times 10^{-3}
\label{fracdif_alpha_hc}
\eeq
\beq
\frac{\alpha_u(sq) - \alpha_\ell(sq)}{\alpha_{ave}(sq)} = 1.96 \times 10^{-4} 
\label{fracdif_alpha_sq}
\eeq
\beq
\frac{\alpha_u(tri)-\alpha_\ell(tri)}{\alpha_{ave}(tri)}=0.621 \times 10^{-3}
\, 
\label{fracdif_alpha_tri}
\eeq
and for $\alpha_0(\Lambda)$, 
\beq
\frac{\alpha_{0,u}(hc) - \alpha_{0,\ell}(hc)}{\alpha_{0,ave}(hc)} = 
2.57 \times 10^{-2} 
\label{fracdif_alpha0_hc}
\eeq
\beq
\frac{\alpha_{0,u}(sq) - \alpha_{0,\ell}(sq)}{\alpha_{0,ave}(sq)} = 
1.13 \times 10^{-2} 
\label{fracdif_alpha0_sq}
\eeq
\beq
\frac{\alpha_{0,u}(tri) - \alpha_{0,\ell}(tri)}{\alpha_{0,ave}(tri)} = 
1.15 \times 10^{-2} 
\label{fracdif_alpha0_tri}
\eeq

Since our lower and upper bounds are so close to each other, we can use them to
obtain an approximate ($ap$) value of the given exponential growth
constant. One way to get this is simply to use the average of the
lower and upper bounds for each exponential growth constant, 
\beq
EGC_{ap}(\Lambda) = EGC_{ave}(\Lambda) \pm \delta_{EGC(\Lambda)}
\label{egc_value}
\eeq
where the uncertainty $\delta_{EGC(\Lambda)}$ is defined as 
\beq
\delta_{EGC(\Lambda)} = (EGF)_u(\Lambda) - EGC_{ave}(\Lambda) = 
EGC_{ave}(\Lambda) - (EGF)_\ell(\Lambda)
\label{egc_delta}
\eeq
Carrying out this procedure, we obtain the following approximate values, 
again listed in order of increasing vertex degree: 
\beq
\alpha_{ap}(hc) = 2.78284 \pm 
                   0.00064 
\label{alpha_hc_average}
\eeq
\beq
\alpha_{ap}(sq) = 3.49359 \pm 
                   0.00034
\label{alpha_sq_average}
\eeq
\beq
\alpha_{ap}(tri) = 4.4733 \pm 
                    0.0014
\label{alpha_tri_average}
\eeq
\beq
\alpha_{0,ap}(hc) = 2.134 \pm 
                     0.027 
\label{alpha0_hc_average}
\eeq
\beq
\alpha_{0,ap}(sq) = 2.846 \pm 
                     0.016
\label{alpha0_sq_average}
\eeq
\beq
\alpha_{0,ap}(tri) = 3.7589 \pm 
                      0.0215 \ . 
\label{alpha0_tri_average}
\eeq
These values are listed in Table \ref{egc_values_table}. As is evident from
these results, we achieve high accuracies in the determinations of these
exponential growth constants, with fractional uncertainties ranging from 
$O(10^{-4})$ to $O(10^{-2})$. 

For each exponential growth constant $EGC(\Lambda)$ that is not known exactly,
we define the estimated ratio from Eq. (\ref{rho}) as
\beq
\rho_{EGC,ap}(\Lambda) \equiv \frac{EGC_{ave}(\Lambda)}{\epsilon(\Lambda)} \ .
\label{rhoest}
\eeq
Regarding the EGCs $\alpha(tri)$, $\alpha_0(tri)$, and $\beta(hc)$, for
which we have presented exact values, we define $\rho_{EGC}(\Lambda)$ as these
exact EGCs divided by $\epsilon(\Lambda)$ for the given lattice, i.e., 
$\rho_{\alpha}(tri) = \alpha(tri)/\epsilon(tri)$, etc. We list these in Table
\ref{rho_values_table}.  

Combining our calculations of lower and upper bounds for the triangular 
lattice with our exact results for this lattice yields another demonstration of
the very high precision of our bounds.  The fractional difference between our
upper bounds and the exact values of $\alpha(tri)$ and $\alpha_0(tri)$ are
extremely small: 
\beq
\frac{\alpha_u(tri)-\alpha(tri)}{\alpha(tri)} = 0.663 \times 10^{-5}
\label{upper_versus_exact_alpha_tri}
\eeq
and
\beq
\frac{\alpha_{0,u}(tri)-\alpha_0(tri)}{\alpha_0(tri)} = 2.53 \times 10^{-3} 
\ . 
\label{upper_versus_exact_alpha0_tri}
\eeq
The corresponding fractional differences relative to our lower bounds are
\beq
\frac{\alpha(tri)-\alpha_\ell(tri)}{\alpha(tri)} = 0.614 \times 10^{-3} 
\label{lower_versus_exact_alpha_tri}
\eeq
and
\beq
\frac{\alpha_0(tri)-\alpha_{0,\ell}(tri)}{\alpha_0(tri)} = 
0.890 \times 10^{-2}  \ . 
\label{lower_versus_exact_alpha0_tri}
\eeq
Thus, our upper bounds for $\alpha(tri)$ and $\alpha_0(tri)$ are closer to the
respective exact results than are the lower bounds.  Consequently, the average
quantities $\alpha_{ave}(tri)$ and quantities $\alpha_{0,ave}(tri)$ lie
slightly below the respective exact values: 
\beq
\frac{\alpha(tri)-\alpha_{ave}(tri)}{\alpha(tri)} = 3.04 \times 10^{-4} 
\label{exact_versus_ave_alpha_tri}
\eeq
and
\beq
\frac{\alpha_0(tri)-\alpha_{0,ave}(tri)}{\alpha_0(tri)} = 3.18 \times 10^{-3} 
\ . 
\label{exact_versus_ave_tri}
\eeq
Provided that this pattern also holds for the square and honeycomb lattices,
then the average quantities $\alpha_{ave}(\Lambda)$ and
$\alpha_{0,ave}(\Lambda)$ and thus the central values of
$\alpha_{ap}(\Lambda)$ and $\alpha_{0,ap}(\Lambda)$ would also lie slightly
below the respective exact values $\alpha(\Lambda)$ and $\alpha_0(\Lambda)$ for
$\Lambda=sq, \ hc$. 

We have also carried out corresponding calculations of lower and upper bounds
for a nonplanar lattice denoted $sq_d$.  We described above how one constructs
a strip of this lattice.  The construction here is analogous. One starts with
the square lattice and then adds (i) an edge connecting the upper left and
lower right vertices of each square to each other, and (ii) an edge connecting
the upper right and lower left vertices of each square to each other.  A finite
section of this lattice with doubly periodic boundary conditions is
a $\Delta$-regular lattice graph with $\Delta_{sq_d}=8$, and this also
describes the infinite planar lattice. We present the
resultant lower bounds and their ratios for $\alpha(sq_d)$ and for
$\alpha_0(sq_d)$ in Tables
\ref{lowerbounds_alpha_sqd_table}-\ref{lowerbounds_alpha0_sqd_table}.  For
upper bounds on this lattice, our relevant results are, first, that the ratio
$\lambda_{sq_d, L_y+1, free}(-1)/\lambda_{sq_d,L_y,free}(-1)$ takes the
respective values 6, $3+(1/3)\sqrt{69}=5.76887462$, and 5.75046353 for
$L_y=1,2,3$, respectively. For the quantity $R_{sq_d,
  \frac{L_y^2}{(L_y-1)(L_y+1)}, free}(-1)$, the indicated pairs yield the
respective values 1.04006421 and 1.00320167. Second, we find that the ratio
$\lambda_{sq_d, L_y+1, free} (0) / \lambda_{sq_d, L_y, free} (0)$ has the
values 6, $(11+\sqrt{97})/4=5.21221445$, and 5.12783026 for $L_y=1,2,3$. For
the quantity $R_{sq_d, \frac{L_y^2}{(L_y-1)(L_y+1)}, free} (0)$, the indicated
pairs yield the respective values 1.15114220 and 1.01645612.  Our results for
the $sq_d$ lattice are
\beq
5.354782509 < \alpha(sq_d) < 
5.750463529
\label{alpha_bounds_sqd}
\eeq
and
\beq
4.417285760 < \alpha_0(sq_d) < 
5.127830256 \ . 
\label{alpha0_bounds_sqd}
\eeq
These bounds for the $sq_d$ lattice (with $\Delta(sq_d)=8$) are included mainly
for the general insight that they yield concerning the dependence of the
exponential growth constants on vertex degree. This goal is already achieved
with the widths that we have included, showing that for the set of honeycomb,
square, triangular, and $sq_d$ lattices $\Lambda$, the quantities
$\alpha(\Lambda)$ and $\alpha_0(\Lambda)$ are monotonically increasing
functions of the vertex degree, $\Delta(\Lambda)$. Accordingly, we have not
attempted to carry out calculations on wider strips of the $sq_d$ lattice to
obtain the same precision in the lower and upper bounds that we did for the
planar lattices considered here, and the bounds (\ref{alpha_bounds_sqd}) and
(\ref{alpha0_bounds_sqd}) are not as restrictive as the corresponding bounds
for the other lattices considered here. 

Given the rapid convergence of our results for these exponential growth
constants on these lattice strips, even for modest strip widths, one could use
extrapolation techniques to infer the actual respective values for $L_y \to
\infty$ in the cases where exact results are not known. However, this
extrapolation analysis is beyond the scope of our present paper, since the
estimation of the uncertainty in the inferred value of $\alpha(\Lambda)$ and
$\alpha_0(\Lambda)$ would depend on the extrapolation method used. These
comments also apply to our bounds on $\beta(\Lambda)$ to be presented below.
It is straightforward to calculate corresponding lower and upper bounds for the
quantities $\rho_\alpha(\Lambda)$ and $\rho_{\alpha_0}(\Lambda)$ for these
lattices; we do not list these explicitly.


\section{Lower and Upper Bounds for $\beta(\Lambda)$}
\label{beta_values_section} 

Using results on calculations of Potts/Tutte polynomials for a variety of
families of lattice strip graphs, we have also obtained lower and upper bounds
on the exponential growth constant $\beta$ for totally cyclic orientations on
these lattices.  As discussed above, these involve dominant $\lambda$ functions
evaluated at $(q,v)=(-1,1)$ or equivalently, $(x,y)=(0,2)$.  
We recall the lower and upper bounds that we have inferred in
Eqs. (\ref{beta_lowerbound}) and (\ref{beta_upperbound}). 

We list our numerical values of lower and upper
bounds for $\beta(\Lambda)$ on various two-dimensional lattices in 
Tables \ref{lowerbounds_beta_sq_table}-\ref{upperbounds_beta_hc_table}. 
The format of these tables is analogous to the format in the
corresponding tables presented above for $\alpha$ and $\alpha_0$.

The relation (\ref{betahc_alphatri}) enables us to obtain a precise value of 
$\beta(hc)$ from our evaluation of the exact expression for 
$\alpha(tri)$ in \cite{baxter87}.  We find 
\beq
\beta(hc) = \sqrt{\alpha(tri)} = 2.11533621655 \ .
\label{beta_hc}
\eeq
Since $\epsilon(hc)=2^{3/2}$, it follows that
\beq
\rho_\beta(hc) = 0.7478842916 \ . 
\label{rho_beta_hc}
\eeq

Summarizing the lower and upper bounds for the these lattices $\Lambda$ from
the above calculations, listed in order of increasing (uniform) vertex 
degree, $\Delta(\Lambda)$, we have
\beq
(*) \quad 2.09444676 < \beta(hc) < 2.12591038 
\label{beta_bounds_hc}
\eeq
\beq
(*) \quad 3.449673447 < \beta(sq) < 3.535730951 
\label{beta_bounds_sq}
\eeq
and
\beq
(*) \quad 7.696127303 < \beta(tri) < 7.832553170  \ . 
\label{beta_bounds_tri}
\eeq
where $(*)$ means that by using duality and our previous calculations of lower
and upper bounds on $\alpha(\Lambda)$ for $\Lambda=sq, \ tri, \ hc$, we can
improve upon these bounds.  Thus, first, using Eq. (\ref{betahc_alphatri}), we
improve upon the bounds (\ref{beta_bounds_hc}):
\beq
2.114686349 < \beta(hc) < 2.115343229 \ .
\label{best_beta_bounds_hc}
\eeq
Evidently, these lower and upper bounds on $\beta(hc)$ are very close to the
precise value, $\beta(hc)=2.11533621655$ in (\ref{beta_hc}).  Second,
using Eq. (\ref{alfbet_sq}) in conjunction with our bounds
(\ref{alpha_bounds_sq}), we improve upon the bounds (\ref{beta_bounds_sq}):
\beq
3.493244874 < \beta(sq) < 
3.493927960 \ .
\label{best_beta_bounds_sq}
\eeq
Third, using Eq. (\ref{betatri_alphahc}), we improve upon the bounds 
(\ref{beta_bounds_tri}): 
\beq
7.740620193 < \beta(tri) < 
7.747796928 \ .
\label{best_beta_bounds_tri}
\eeq
Aside from $\beta(hc)$, for which we have given an exact value, these lower and
upper bounds on $\beta(\Lambda)$, (\ref{best_beta_bounds_sq}) and
(\ref{best_beta_bounds_tri}) are, to our knowledge, the best current bounds on
these quantities.  

As was the case with our lower and upper bounds on $\alpha(\Lambda)$ and
$\alpha_0(\Lambda)$, our lower and upper bounds on $\beta(\Lambda)$ are very
close to each other. To show this, we exhibit the fractional differences of the
upper and lower bounds on these lattices, in order of increasing vertex degree:
\beq
\frac{\beta_u(hc) - \beta_\ell(hc)}{\beta_{ave}(hc)} = 3.11 \times 10^{-4} 
\label{fracdif_beta_hc}
\eeq
\beq
\frac{\beta_u(sq) - \beta_\ell(sq)}{\beta_{ave}(sq)} = 1.96 \times 10^{-4} 
\label{fracdif_beta_sq}
\eeq
(which is the same as Eq. (\ref{fracdif_alpha_sq}) by duality) and
\beq
\frac{\beta_u(tri) - \beta_\ell(tri)}{\beta_{ave}(tri)} = 0.927 \times 10^{-3}
  \ . 
\label{fracdif_beta_tri}
\eeq

Hence, as before, since our lower and upper bounds are quite close to each 
other, we can use them to obtain the approximate value of
$\beta(\Lambda)$ on the various lattices $\Lambda$. Using the same procedure as
dicussed above in Eqs. (\ref{egc_average}), (\ref{egc_value}), and 
(\ref{egc_delta}), we calculate the approximate values 
\beq
\beta_{ap}(hc) = 2.11501 \pm 
                  0.00033
\label{beta_hc_average}
\eeq
\beq
\beta_{ap}(sq) = 3.49359 \pm 
                  0.00034  
\label{beta_sq_average}
\eeq
(which is the same as Eq. (\ref{alpha_sq_average}) by duality) and
\beq
\beta_{ap}(tri) =  7.7442 \pm 
                    0.0036 \ . 
\label{beta_tri_average}
\eeq
These values of $\beta_{ap}(sq)$ and $\beta_{ap}(tri)$ are listed in Table
\ref{egc_values_table}, which also includes our exact value for $\beta(hc)$.
As is evident, we achieve very high accuracy with our determination of these
approximate values, with a fractional uncertainty of less than $10^{-4}$ for 
$\beta(sq)$ and $5 \times 10^{-4}$ for $\beta(tri)$.

In Table \ref{rho_values_table} we list the values of the ratios 
$\rho_\alpha(\Lambda)$,
$\rho_{\alpha_0}(\Lambda)$, and $\rho_\beta(\Lambda)$ obtained from our
calculations. For $\rho_\alpha(tri)$, $\rho_{\alpha_0}(tri)$, and 
$\rho_\beta(hc)$, we list the exact values, and for the others we list the
ratios calculated using $EGC_{ave}(\Lambda)$, where $EGC=\alpha, \ \alpha_0, \
\beta$. 

We can use our exact value of $\beta(hc)$ to obtain a further measure of the
accuracy of our bounds.  The fractional differences between our upper and lower
bounds on $\beta(hc)$ and this exact value are 
\beq
\frac{\beta_u(hc)-\beta(hc)}{\beta(hc)} = 3.31 \times 10^{-6} 
\label{upper_versus_exact_beta_hc}
\eeq
and
\beq
\frac{\beta(hc)-\beta_\ell(hc)}{\beta(hc)} = 3.07 \times 10^{-4}  \ . 
\label{exact_versus_lower_beta_hc}
\eeq
Thus, as was true of our lower and upper bounds on $\alpha(tri)$ and
$\alpha_0(tri)$, here we observe that the upper bound in
(\ref{best_beta_bounds_hc}) is closer to the exact value, $\beta(hc)$ than is
the lower bound in (\ref{best_beta_bounds_hc}).  Hence, the average,
$\beta_{ave}(hc)$ is slightly below the exact value:
\beq
\frac{\beta(hc)-\beta_{ave}(hc)}{\beta(hc)} = 1.52 \times 10^{-4} \ . 
\label{exact_versus_ave_beta_hc}
\eeq
As before, if this pattern also holds for the square and triangular lattices,
then the average quantities $\beta_{ave}(\Lambda)$ and
thus the central values of $\beta_{ap}(\Lambda)$ would also be slightly
smaller than the exact values $\beta(\Lambda)$ for $\Lambda=sq, \ tri$. 

As discussed above, we have included results for the $sq_d$ 
lattice here for the information that they give on the dependence of the
exponential growth constants on vertex degree. 
For the $sq_d$ lattice strip with cyclic BCs and $L_y=2$, we calculate 
\beq
[\lambda_{sq_d, L_y, cyc}(-1,1)]^{1/2} = \sqrt{13+\sqrt{181}} = 5.14330867
\label{lamcyc_sqd}
\eeq
and for toroidal BCs and $L_y=3$, we obtain the numerical value 
\beq
[\lambda_{sq_d, L_y, tor}(-1,1)]^{1/2} = 15.85636130 
\ .  
\label{lamtor_sqd}
\eeq
These yield lower bounds on $\beta(sq_d)$.  In addition, we calculate the ratio
\beq
\frac{\lambda_{sq_d, 2, cyc} (-1,1)}{\lambda_{sq_d, 1 cyc} (-1,1)} = 
13+\sqrt{181} = 26.45362405 \ , 
\label{lamrat_sqd}
\eeq
which yields an upper bound on $\beta(sq_d)$. Thus, we have the loose bounds
\beq
15.8563613 < \beta(sq_d) < 26.4536240 \ . 
\label{beta_bounds_sqd}
\eeq
These bounds are much less stringent than our bounds on $\beta(\Lambda)$ for
the other lattices, but they are sufficient to show the monotonic increase of
$\beta(\Lambda)$ with vertex degree $\Delta(\Lambda)$ among these lattices. For
this reason, we have not tried to include results from larger-width strips for
this $sq_d$ lattice. 


\section{Comparative Analysis}
\label{comparison_section}


\subsection{General} 

From our calculations, we observe that for these lattices $\Lambda$, the values
of $\alpha(\Lambda)$, $\alpha_0(\Lambda)$, and $\beta(\Lambda)$ that are
consistent with our lower and upper bounds (and the exact values where we have
calculated them) are monotonically increasing functions of $\Delta(\Lambda)$.
In particular, this is true of the quantities $\alpha_{ave}(\Lambda)$,
$\alpha_{0,ave}(\Lambda)$, and $\beta_{ave}(\Lambda)$.  This is the opposite of
the behavior of the ground-state degeneracy of the $q$-state Potts
antiferromagnet, $W(\Lambda,q)$, which, for values of $q$ used in proper
$q$-colorings of $\Lambda$, is a monotonically decreasing function of
$\Delta(\Lambda)$. This dependence of $W(\Lambda,q)$ on $\Delta(\Lambda)$ was
also shown in lower and upper bounds on $W(\Lambda,q)$ \cite{w},
\cite{ww}-\cite{ilb}. The fact that, for a given value of $q$ used for a proper
$q$-coloring of the lattice $\Lambda$, $W(\Lambda,q)$ is a monotonically
decreasing function of $\Delta(\Lambda)$, was shown to be a consequence of the
fact that increasing $\Delta(\Lambda)$ places more constraints on this proper
$q$-coloring \cite{w,bcc99,wn}. The reversal in the dependence of
$W(\Lambda,q)$ on $\Delta(\Lambda)$ going from (positive) values of $q$ used in
proper $q$-colorings of $\Lambda$ to $q \le 0$ was evident in (Fig. 5 of)
Ref. \cite{w}. Our present results extend these earlier ones with quite
restrictive upper and lower bounds and high-accuracy approximate values for
$\alpha(\Lambda) = |W(\Lambda,-1)|$ and $\alpha_0(\Lambda) = |W(\Lambda,0)|$. A
property that is pertinent here is the fact that the signs of successive terms
in the chromatic polynomial alternate, starting with a positive sign (and,
indeed, a coefficient of unity) for the highest-degree term, $q^{n(G)}$, then a
negative sign for the $q^{n(G)-1}$ term, and so forth for lower-power
terms. Hence, if $q$ is positive, as in the evaluation of $W(G,q)$ for the
ground state degeneracy of the $q$-state Potts antiferromagnet, then alternate
terms contribute with opposite sign, whereas if $q$ is negative, as in the
evaluations at $q=-1$ for $a(G)$ and $a_0(G)$, then all of the terms in
$P(G,q)$ and $P_r(G,q)=q^{-1}P(G,q)$ contribute with the same sign.

As regards the relative sizes of $\alpha(\Lambda)$ and $\alpha_0(\Lambda)$, on
the one hand, and $\beta(\Lambda)$ on the other, we find that $\alpha(\Lambda)$
and $\alpha_0(\Lambda)$ may be larger or smaller than $\beta(\Lambda)$, while
the duality of the square lattice implies the equality of $\alpha(sq)$ and
$\beta(sq)$.  We also observe that the property that two families of recursive
lattice graphs have the same value of $\Delta$ (or $\Delta_{eff}$) does not
imply that they have the same values of $\alpha(\{G\})$, $\alpha_0(\{G\})$, or
$\beta(\{ G\})$.  For example, the cyclic square-lattice strip graph $L_m$ has
the same value of (uniform) vertex degree $\Delta=3$ as the honeycomb lattice,
but the values of $\alpha(\{L\})$, $\alpha_0(\{L\})$, and $\beta(\{L\})$ are
different from the respective values of $\alpha(hc)$, $\alpha_0(hc)$, and
$\beta(hc)$.

We recall the inequality (\ref{alpha0_lt_alpha}). From our bounds and exact
results, we compute the ratios $\alpha_0(\Lambda)/\alpha(\Lambda)$ for various
lattices $\Lambda$, using $\alpha_{ave}(\Lambda)$ and $\alpha_{0,ave}(\Lambda)$
for $\Lambda=hc, \ sq$ and our exact values $\alpha(tri)$ and
$\alpha_0(tri)$. In order of increasing vertex degree, we have
\beq
\frac{\alpha_{0,ave}(hc)}{\alpha_{ave}(hc)} = 0.767
\label{alpha0_over_alpha_hc}
\eeq
\beq
\frac{\alpha_{0,ave}(sq)}{\alpha_{ave}(sq)} = 0.815
\label{alpha0_over_alpha_sq}
\eeq
and
\beq
\frac{\alpha_0(tri)}{\alpha(tri)} = 0.8427300 \ . 
\label{alpha0_over_alpha_tri}
\eeq
We note that for these lattices, this ratio is a monotonically increasing
function of $\Delta(\Lambda)$. This is the same dependence that we showed for
the infinite-length, finite-width strips discussed in Section
\ref{example_section}. 

We have also obtained results on these exponential growth constants for a
number of heteropolygonal Archimedean lattices (i.e., Archimedean lattices
comprised of more than one type of regular polygon). These will be reported
elsewhere \cite{aca}. 

Concerning the ratios $\rho_{EGC}(\Lambda)$, we find that 
$\rho_\alpha(\Lambda)$ and $\rho_{\alpha_0}(\Lambda)$ are monotonically 
decreasing functions, while $\rho_\beta(\Lambda)$ is a monotonically 
increasing function of $\Delta(\Lambda)$.  Again, this is the same dependence
that we found for infinite-length, finite-width strips as a function of 
$\Delta(\{ G \})$ (or, where appropriate, $\Delta_{eff} ( \{ G \})$). 

Using similar methods, we have also obtained results on exponential growth
constants for spanning forests and connected spanning subgraphs on a variety of
lattices. A spanning forest in a graph $G$ is a spanning subgraph of $G$ that
does not contain any circuits. Denote $N_{SF}(G)$ as the number of spanning
forests of a graph $G$ and $\phi(\{G\}) \equiv \lim_{n(G) \to \infty}
[N_{SF}(G)]^{1/n(G)}$. For example, for the square lattice, we have found
$3.675183 \le \phi(sq) < 3.699659$, improving on the bounds $3.32 \le \phi(sq)
\le 3.8416195$ in \cite{merino_welsh99}, the bounds $3.64497 \le \phi(sq) \le
3.74101$ in \cite{cmnn}, the bounds $3.65166 \le \phi(sq) \le 3.73635$ in
\cite{garijo2014}, and the upper bound $\phi(sq) \le 3.705603$ in
\cite{mani2012}. For the triangular and honeycomb lattices we obtain $5.393333
\le \phi(tri) \le 5.494840$ and $2.803787 \le \phi(hc) \le 2.804781$. Details
will be reported elsewhere \cite{aca}.


\subsection{Comparison with Spanning Trees}

The quantities $a(G)$, $a_0(G)$, and $b(G)$ enumerate classes of orientations
of arrows defined on the edges of the directed graph $D(G)$, but depend only on
$G$ itself, and similarly with the resultant exponential growth constants in
the $n(G) \to \infty$ limit.  Because of this, it is appropriate to compare
them with the number of spanning trees on $G$ and the associated exponential
growth constant.  We do this in the present section. Recall that a tree graph
is defined as a connected graph that does not contain any circuits, and a
spanning tree of a graph $G$ is a subgraph of $G$ that is a tree and that
contains all of the vertices of $G$ (and a subset of the edges of $G$). 

From the definition (\ref{t}), it is evident that the number of spanning trees
in a graph $G$ is 
\beq
N_{ST}(G)=T(G,1,1) \ . 
\label{nst}
\eeq
Since the nonzero coefficients of each term in Eq. (\ref{tij}) are positive,
and since $a_0(G)=T(G,1,0)$ (recall Eq. (\ref{a0_tx1y0})), a basic inequality
is \cite{merino_welsh99} 
\beq
a_0(G) \le N_{ST}(G) \ , \quad i.e., \quad  T(G,1,0) \le T(G,1,1) \ . 
\label{a0_lt_nst}
\eeq
The necessary and sufficient condition for this to be an equality is clear from
Eqs. (\ref{a0_tx1y0}), (\ref{nst}), and (\ref{tij}); thus, $T(G,1,0)=T(G,1,1)$
if and only if $T(G,x,y)$ does not contain any nonzero terms of the form 
$t_{ij} x^i y^j$ with $j \ge 1$. From the definition (\ref{t}), this condition
is equivalent to the condition that $G$ does not contain any cycles. 

For the families of graphs under consideration here, $N_{ST}(G)$ grows
exponentially rapidly with the number of vertices, $n(G)$.  This motivates one
to analyze the associated exponential growth constant,
\beq
\tau(\{G\}) \equiv \lim_{n(G) \to \infty} [N_{ST}(G)]^{1/n(G)} \ . 
\label{tau} 
\eeq
An equivalent quantity is $z(\{G\})$, defined as 
\beq
z(\{G\}) \equiv \ln[\tau(\{ G \})] \ . 
\label{zg}
\eeq
The exponential growth constants $\tau(\Lambda)$ have been calculated exactly
for the square, triangular, and honeycomb lattices under consideration here
\cite{wu77}, and, indeed, for all Archimedean lattices \cite{st,std,sti} (as
well as some higher-dimensional lattices). The relevant exponential growth
constants for the planar lattices studied here are, in order of increasing
vertex degree, \cite{wu77}
\beq
\tau(hc) = \exp \Big [ \frac{\ln(3)}{4} + 
\frac{3}{\pi} \, {\rm Ti}_2 \Big (\frac{1}{\sqrt{3}} \Big ) \Big ] = 
2.24266494889
\label{tau_hc}
\eeq
\beq
\tau(sq) = \exp \Big ( \frac{4C}{\pi} \Big ) = 3.20991230073
\label{tau_sq} 
\eeq
and
\beq
\tau(tri) = \exp \bigg [ \frac{\ln(3)}{2} + 
\frac{6}{\pi} \, {\rm Ti}_2 \Big (\frac{1}{\sqrt{3}} \Big ) \bigg ] 
= 5.02954607297 \ . 
\label{tau_tri}
\eeq
(to the indicated precision).  In Eqs. (\ref{tau_hc}) and (\ref{tau_tri}),
Ti$_2(x)$ is the tangent inverse integral,
\beq
{\rm Ti}_2(x) = \int_0^{x} \frac{{\rm arctan}(y)}{y} \, dy 
        = \sum_{n=0}^\infty \frac{(-1)^n \, x^{2n+1}}{(2n+1)^2} \ , 
\label{tifun}
\eeq
and in Eq. (\ref{tau_sq}), $C$ is the Catalan constant, 
\beq
C = \sum_{n=0}^\infty \frac{(-1)^n}{(2n+1)^2} = {\rm Ti}_2(1) = 
0.915965594177
\label{catalan}
\eeq 
Owing to the fact that the triangular
lattice is the (planar) dual of the honeycomb lattice and using 
Eq. (\ref{nuhctri}), we have 
\beq
\tau(hc) = [\tau(tri)]^{1/2} \ , 
\label{tau_tri_hc_rel}
\eeq
which is evident in Eqs. (\ref{tau_hc}) and (\ref{tau_tri}). 
For the $sq_d$ lattice, we have \cite{sti}
\beq
\tau(sq_d) = \exp \Bigg [ \frac{4C}{\pi} +
\ln(2-\sqrt{3}) - \frac{4}{3} {\rm arctanh} \Big (\frac{1}{\sqrt{3}} \Big ) +
\frac{4}{\pi} {\rm Ti}_2(2+\sqrt{3}) \Bigg ] = 6.984820959 \ .
\label{tau_sqd}
\eeq

For the strip graphs of the honeycomb, square, and triangular lattices
discussed in Section \ref{example_section}, the exponential growth constants 
$\tau$ are, again in order of increasing $\Delta$ or $\Delta_{eff}$
\cite{a,ta,hca,ka,sdg}, 
\beq
\tau(\{ HL \}) \equiv \tau( hc, 2_F \times \infty) = 
(3 + 2\sqrt{2} \, )^{1/4} = \sqrt{1+\sqrt{2}} = 1.55377397
\label{tau_hc_strip}
\eeq
\beq
\tau(\{ L \}) \equiv \tau(sq,2_F \times \infty) = \sqrt{2+\sqrt{3}} = 
1.93185165
\label{tau_sq_strip}
\eeq
\beq
\tau( \{ Wh \}) = \frac{3+\sqrt{5}}{2} = 2.61803399
\label{tau_wh_strip}
\eeq
\beq
\tau( \{ TL \}) \equiv \tau(tri,2_F \times \infty) = 
\sqrt{ \frac{7+3\sqrt{5}}{2} \,} = 
\frac{3+\sqrt{5}}{2} = 2.61803399 
\label{tau_tri_strip}
\eeq
and
\beq
\tau( \{ sq_d \}) \equiv \tau((sq_d)_{2,F} \times \infty) = 2\sqrt{3} = 
3.4641016 \ . 
\label{tau_sqd_strip}
\eeq
(Note that $\alpha(\{sq_d\})=\tau(\{sq_d\})$ in the case of the $sq_d$ strip.)
For both the infinite planar lattices and for these infinite-length,
finite-width lattice strip graphs, the exponential growth constant $\tau(\{ G
\})$ is a monotonically increasing function of $\Delta(\Lambda)$ (and, where
appropriate, $\Delta_{eff}(\{ G \})$). 

We discuss some inequalities.  A theorem of Thomassen \cite{thomassen2010}
states that if $G$ is a $\Delta$-regular graph of degree $\Delta(G) \le 3$
which has no loops (which may have bridges and multiple edges), then $N_{ST}(G)
\le a(G)$.  Considering a family of graphs of this type and taking the limit
$n(G) \to \infty$, this implies that in this limit, $\tau(\{G\}) \le \alpha(\{
G \})$.  It is readily checked that our results for $\Delta$-regular families
of graphs of degree $\Delta(G) \le 3$ satisfy this theorem. For example, for
the cyclic square-ladder strip $L_m$, which has $\Delta(L_m)=3$, the value of
$\tau(\{L\})$ given in Eq.  (\ref{tau_sq_strip}), namely $\tau(\{L\})=1.932$,
is less than $\alpha(\{L\})=\sqrt{7}=2.646$ given in Eq. (\ref{alpha_sqlad}),
and for the infinite honeycomb lattice, with $\Delta(hc)=3$, the exactly known
value of $\tau(hc)$ given in Eq. (\ref{tau_hc}) from \cite{wu77}, is less than
our value for $\alpha(hc)$ given in Eq. (\ref{alpha_hc_average}), namely
$\alpha_{ap}(hc) = 2.78284 \pm 0.00064$.  We calculate the ratio
\beq
\frac{\tau(hc)}{\alpha_{ap}(hc)} = 0.80589 \pm 0.00019 \ . 
\label{tau_hc_over_alpha_hc}
\eeq

Another theorem of Thomassen \cite{thomassen2010} states that if $G$ is a
graph with no loops or bridges (but which may have multiple edges) with 
$e(G) \ge 4[n(G)-1]$, then
$N_{ST}(G) \le b(G)$. For a family of graphs satisfying this condition, in the
limit $n(G) \to \infty$, this theorem implies that 
$\tau(\{G\}) \le \beta(\{G\})$.  
Among the graphs that we consider, sections of the
$sq_d$ lattice with doubly periodic boundary conditions, and also the infinite
$sq_d$ lattice have $\Delta(sq_d)=8$ and hence $e(G)=4n(G)$. Again, it is
readily checked that, with the value of $\tau(sq_d)=6.985$ in
Eq. (\ref{tau_sqd}), it follows that any value of 
$\beta(sq_d)$ in the range allowed by our inferred lower and upper bounds in 
(\ref{beta_bounds_sqd}) satifies this theorem.

For all of the lattice graphs that we consider, 
\beq
\alpha_0( \{ G \}) < \tau( \{ G \}) \ .
\label{alpha0_lt_tau}
\eeq
Note that this inequality is not implied by the inequality 
(\ref{a0_lt_nst}), since, {\it a priori}, the difference, 
$\lim_{n(G) \to \infty} [N_{ST}]^{1/n(G)} - 
 \lim_{n(G) \to \infty} [a_0(G)]^{1/n(G)}$ 
might vanish as $n(G) \to \infty$. 

Furthermore, given the lower and upper bounds (\ref{alpha_bounds_sq}) and the
duality relation that $\alpha(sq)=\beta(sq)$, it is evident that
\beq
\alpha(sq) > \tau(sq) \ .
\label{alpha_sq_gt_tau_sq}
\eeq

Numerically, using our determination of the approximate value 
$\alpha_{ap}(sq)$ in Eq. (\ref{alpha_sq_average}), we have
\beq
\frac{\tau(sq)}{\alpha_{ap}(sq)} = \frac{\tau(sq)}{\beta_{ap}(sq)} = 
0.9188005 \pm 0.0000894 \ . 
\label{tau_sq_over_alpha_sq}
\eeq
From the exact value of $\alpha(tri)$ in Eq. (\ref{alpha_tri}), we have
\beq
\alpha(tri) < \tau(tri) \ . 
\label{alpha_tri_lt_tau_tri}
\eeq
Numerically, 
\beq
\frac{\alpha(tri)}{\tau(tri)} = 0.8896722 \ . 
\label{alpha_tri_over_tau_tri}
\eeq
Given our bounds on $\beta(tri)$, (\ref{best_beta_bounds_tri}), we have
\beq
\beta(tri) > \tau(tri) \ . 
\label{beta_tri_gt_tau_tri}
\eeq
Using our determination of $\beta_{ap}(tri)$ in Eq. (\ref{beta_tri_average}),
we compute 
\beq
\frac{\tau(tri)}{\beta_{ap}(tri)} = 0.64946 \pm 0.00030 \ . 
\label{tau_tri_over_beta_tri}
\eeq

Finally, from the exact value (\ref{beta_hc}), we have the inequality 
\beq
\beta(hc) < \tau(hc) 
\label{beta_hc_lt_tau_hc}
\eeq
and numerically, 
\beq
\frac{\beta(hc)}{\tau(hc)} = 0.943224 \ . 
\label{beta_hc_over_tau_hc}
\eeq

It is of interest to discuss how our results relate to the Merino-Welsh
conjecture (MWC) \cite{merino_welsh99} and a later conjecture by Conde and
Merino (CMC) \cite{conde_merino2009}.  The Merino-Welsh conjecture is as
follows: Let $G$ be a connected graph without loops or bridges (which may
have multiple edges). Then the Merino-Welsh conjecture is the inequality
\cite{merino_welsh99}
\beq
N_{ST}(G) \le {\rm max}[a(G), \ b(G)] \ , \quad i.e., \ 
T(G,1,1) \le {\rm max}[T(G,2,0), \ T(G,0,2)] \quad {\rm (MWC)} \ .
\label{mwc} 
\eeq
Subsequently, in \cite{conde_merino2009}, Conde and Merino conjectured the
stronger inequality that if $G$ is a connected graph without loops or bridges
(which may have multiple edges), then
\beq
[N_{ST}(G)]^2 \le a(G)b(G) \ , \quad i.e., \ 
[T(G,1,1)]^2 \le T(G,2,0)T(G,0,2) \quad {\rm (CMC)} \ .
\label{cmc} 
\eeq
As observed in \cite{conde_merino2009}, the inequality (\ref{cmc}) implies the
inequality (\ref{mwc}).  Some works related to these conjectures include
\cite{merino_note2009}-\cite{knauer2016}. In
particular, the Merino-Welsh conjecture has been proved for wheel graphs
$Wh_n$, complete graphs $K_n$, and complete bipartite graphs $K_{r,s}$ with $r
\ge s \ge 2$ in \cite{merino_note2009}, and for series-parallel graphs in
\cite{noble_royle2011}.  

We first note that the Merino-Welsh and Conde-Merino conjectures imply
the following inequalities on exponential growth constants: 
\beq
\tau(\{ G \}) \le {\rm max}[ \alpha(\{ G \}), \ \beta( \{ G \})] \quad 
{\rm from \ MWC} \ .  
\label{mwc_egc}
\eeq
and 
\beq
[\tau(\{ G \})]^2 \le \alpha(\{ G \}) \beta( \{ G \}) \quad 
{\rm from \ CMC} \ .  
\label{cmc_egc}
\eeq
We discuss these in turn. 
As is evident in Table \ref{egc_values_table}, our results are in agreement
with the inequality (\ref{mwc_egc}).  This is also true for all of the
infinite-length, finite-width strip graphs discussed in Section
\ref{example_section}, for which the $\tau(\{ G \})$ values were given in 
Eqs. (\ref{tau_hc_strip})-(\ref{tau_sqd_strip}). Recall that the values of
$\tau(\Lambda)$ are exactly known for all of the (infinite limits of) lattice
graphs that we consider here.  For the honeycomb lattice and the cyclic 
square-lattice strip graph, the validity of the
inequality (\ref{mwc_egc}) is guaranteed by the first theorem from Thomassen
\cite{thomassen2010} mentioned above, namely that because $\Delta(hc)=
\Delta(L_m)=3$, it follows that $\tau(hc) \le \alpha(hc)$ and 
$\tau(\{ L\}) \le \alpha(\{ L \})$. In either of the hypothetical cases in
which $\alpha(hc) \ge \beta(hc)$ or $\alpha(hc) \le \beta(hc)$, this implies 
that $\tau(hc) \le {\rm max}[\alpha(hc), \ \beta(hc)]$ and similarly with
$\{ L \}$. In fact, we find that $\alpha(hc) > \beta(hc)$ and 
$\alpha(\{L\}) > \beta(\{L\})$. For the square lattice, we find 
\beq
\tau(sq) < {\rm max}[\alpha(sq), \ \beta(sq)]   \quad {\rm where} \ \ 
\alpha(sq)=\beta(sq), 
\label{mw_sq}
\eeq
with the approximate value of $\tau(sq)/\alpha(sq) = \tau(sq)/\beta(sq)$ 
given by Eq. (\ref{tau_sq_over_alpha_sq}). For the triangular lattice, we have
\beq
\tau(tri) < {\rm max}[\alpha(tri), \ \beta(tri)] = \beta(tri) \ , 
\label{mw_tri}
\eeq
with the approximate value of $\tau(tri)/\beta(tri)$ given by 
Eq. (\ref{tau_tri_over_beta_tri}). 

Our results are also in agreement with the inequality on exponential growth
constants implied by the Conde-Merino conjecture (\ref{cmc_egc}). This is clear
for our results on the infinite-length finite-width strip graphs discussed in
Section \ref{example_section} in conjunction with Eqs.
(\ref{tau_hc_strip})-(\ref{tau_sqd_strip}), all of which are exact. Further,
for the infinite planar lattices we have
\beq
[\tau(hc)]^2 = 5.029546 < \alpha_{ap}(hc)\beta(hc)= 5.8866
\label{cmc_hc}
\eeq
\beq
[\tau(sq)]^2 = 10.30354 < \alpha_{ap}(sq)\beta_{ap}(sq)= 
[\alpha_{ap}(sq)]^2 = 12.205
\label{cmc_sq}
\eeq
and
\beq
[\tau(tri)]^2 = 25.29633 < \alpha(tri)\beta_{ap}(tri) = 34.653 \ ,
\label{cmc_tri}
\eeq
where we have used the approximate ($ap$) values that we have determined for
$\alpha_{ap}(hc)$, $\alpha_{ap}(sq)=\beta_{ap}(sq)$, and $\beta_{ap}(tri)$ in
Eqs. (\ref{alpha_hc_average}), (\ref{alpha_sq_average}), and
(\ref{beta_tri_average}), respectively, together with our exact values for
$\alpha(tri)$ and $\beta(hc)$ in Eq. (\ref{alpha_tri}) and (\ref{beta_hc}), in
computing the right-hand sides of (\ref{cmc_hc})-(\ref{cmc_tri}).  As is clear
from these results, the accuracy with which we have determined the approximate
values $\alpha_{ap}(hc)$, $\alpha_{ap}(sq)=\beta_{ap}(sq)$, and
$\beta_{ap}(tri)$ is more than adequate to establish the validity of the
inequalities (\ref{cmc_hc})-(\ref{cmc_tri}).


\section{Conclusions}
\label{conclusion_section}

In this paper we have calculated the exponential growth constants $\alpha$,
$\alpha_0$, and $\beta$ describing the asymptotic behavior of, respectively,
acyclic orientations, acyclic orientations with a single source vertex, and
totally cyclic orientations of directed lattice strip graphs.  We have
considered several different types of lattices, including square, triangular,
and honeycomb. From our calculations, we have inferred new lower and upper
bounds on these exponential growth constants for the respective infinite
lattices.  Our bounds from calculations on infinite-length, finite-width
lattice strips converge rapidly even for modest values of strip widths. Using
exact results for $\alpha(tri)$, $\alpha_0(tri)$, and $\beta(hc)$, we have
shown that our lower and upper bounds are very close to the exact values of
these quantities. In addition to the above-mentioned exact results, our bounds
are, to our knowledge, the best current bounds on these exponential growth 
constants. Since our
lower and upper bounds are quite close to each other, we infer quite accurate
approximate values for the exponential growth constants that are not exactly
known.  These values have fractional uncertainties ranging from $O(10^{-4})$ to
$O(10^{-2})$. Comparisons of these values with the growth constants for
spanning trees on these lattices are given.  Our results are in agreement with
the Merino-Welsh and Conde-Merino conjectures.  We have also presented
corresponding bounds for a nonplanar lattice denoted $sq_d$ with a higher
vertex degree, $\Delta=8$. Our results show that $\alpha(\Lambda)$,
$\alpha_0(\Lambda)$, and $\beta(\Lambda)$ are monotonically increasing
functions of vertex degree $\Delta(\Lambda)$ for these lattices. We have
conjectured that the analytic expression that was proved to be a lower bound on
$W(\Lambda,q)$ for values of $q$ used in proper $q$-colorings of $\Lambda$ is
an upper bound on $\alpha(\Lambda)$ and $\alpha_0(\Lambda)$ when evaluated at
$q=-1$ and $q=0$, respectively.

The properties that $\alpha(\{G\})$ and $\alpha_0(\{G\})$ involve the
evaluation of $W(\{G\},q)$, the degeneracy, per vertex, of the partition
function of the zero-temperature Potts antiferromagnet, at the respective
values $q=-1$ and $q=0$ on the $n \to \infty$ limit $\{G\}$ of graphs $G$,
while $\beta(\{G\})$ involves the evaluation of the exponent of the
dimensionless free energy, per vertex, of the ferromagnetic Potts model at
$q=-1$ and the finite-temperature value $v=1$, provide a very interesting and
intriguing connection between graph-theoretic quantities and functions in
statistical mechanics. We have taken advantage of this connection in this 
paper. We regard this as a very good
example of the fruitful interplay between mathematics and physics and believe
that further insights will be obtained by exploiting it in the future.


\begin{acknowledgments}

This research was supported in part by the Taiwan Ministry of Science and
Technology grant MOST 103-2918-I-006-016 (S.-C.C.) and by
the U.S. National Science Foundation grant No. NSF-PHY-16-1620628 (R.S.).

\end{acknowledgments}


\begin{appendix}


\section{Some Graph Theory Background}
\label{graphtheory}

In this appendix we include some graph theory background relevant for our
analysis in the paper (for further details, see, e.g., \cite{graphtheory}).  As
in the text, let $G=(V,E)$ be a graph defined by its vertex and edge sets $V$
and $E$.  Let $n=n(G)=|V|$, $e(G)=|E|$, and $k(G)$ denote the number of
vertices, edges, and connected components of $G$, respectively.  Denote
$\Delta(v_i)$ as the degree of the vertex $v_i \in V$.  A loop is defined as an
edge that connects a vertex to itself, and a bridge (co-loop) is defined as an
edge that has the property that if it is deleted, then this increases the
number of components in the resultant graph, relative to the number of
components in the initial graph that contained the bridge. If a graph has no
bridges, then it is said to be 2-connected. Two adjacent vertices may have more
than one edge joining them; if so, one says that the graph has multiple edges
(and sometimes calls it a multigraph). However, as explained in the text, in
order to have minimal measures of acyclic orientations, acyclic orientations
with a unique source, and totally cyclic orientations, we exclude graphs with
loops or multiple edges from our analysis. A cycle on $G$ is defined as a set
of edges that form a closed circuit (cycle). Let $c(G)$ denote the number of
linearly independent cycles in $G$.  This satisfies $c(G)=e(G)+k(G)-n(G)$.  The
``join'' of two graphs $G$ and $H$ is denoted $G+H$ and is defined as the graph
constructed by adding edges to each of the vertices of $G$ connecting to each
of the vertices of $H$.

The chromatic polynomial of $G$, denoted $P(G,q)$, counts the number of ways of
assigning $q$ colors to the vertices of $G$ subject to the condition that no
two adjacent vertices have the same color \cite{graphtheory}-\cite{dkt}.  Such
a color assignment is called a proper $q$-coloring of (the vertices of)
$G$. The chromatic number $\chi(G)$ is defined as the minimum value of $q$
required for a proper $q$-coloring of $G$. A spanning subgraph of $G$, denoted
$G'$, is a graph with the same vertex set $V$ and a subset of the edge set $E$,
i.e., $G'=G'(V,E')$ with $E' \subseteq E$.  $P(G,q)$ is given by
\beq
P(G,q) = \sum_{G' \subseteq G} (-1)^{e(G')} q^{k(G')} \ . 
\label{p}
\eeq
As is clear from Eq. (\ref{p}), $P(G,q)$ is a polynomial of degree $n=n(G)$ in
$q$.  Since one obviously cannot perform a proper $q$-coloring of a graph $G$
if the number of colors is zero, i.e., $q=0$, it follows that $P(G,q)$ always
contains a factor $q$. This property is also clear from Eq. (\ref{p}).
Consequently, one may define a reduced polynomial
\beq
P_r(G,q) \equiv \frac{P(G,q)}{q} \ .
\label{pr}
\eeq
One can write the chromatic polynomial as 
\beq
P(G,q) = \sum_{j=1}^{n(G)} \kappa_j(G) \, q^j \ , 
\label{psum}
\eeq
with $\kappa_n(G)=1$, etc. A general property is that the signs of the
$\kappa_j(G)$ alterate as $j$ decreases from $n$ to 1. From Eq. (\ref{a0_prq0})
and (\ref{psum}), it follows that
\beq
a_0(G) = (-1)^{n(G)-1} \, \kappa_1(G) \ . 
\label{pkappa1}
\eeq

The chromatic polynomial is a special case of the partition function of the
$q$-state Potts model, $Z(G,q,v)$. A convenient expression for this partition 
function is as a sum of contributions from spanning subgraphs \cite{fk}, 
\beq
Z(G,q,v) = \sum_{G' \subseteq G} v^{e(G')} q^{k(G')} \ , 
\label{zcluster}
\eeq
where, in the physics context, $v$ is a temperature-dependent variable given by
Eq. (\ref{veq}) below. As is obvious from Eq. (\ref{zcluster}), $Z(G,q,v)$ is a
polynomial in $q$ and $v$ with the property that the nonzero coefficients are
positive integers. The expression for $Z(G,q,v)$ allows one to define the Potts
partition function for values of $q$ that are more general than just the
positive integers. From (\ref{p}) and (\ref{zcluster}), it is evident that
\beq
P(G,q)=Z(G,q,-1) \ . 
\label{pz}
\eeq

We recall the definition of $Z(G,q,v)$ in terms of a spin-type Hamiltonian
\cite{wurev}.  Consider a graph $G$ with a set of classical spins $\sigma_i$
taking values in the set of positive integers $\{1,...,q\}$ at each site
(vertex) $v_i$ of $G$, whose interactions with spins on adjacent sites are
described by the Hamiltonian
\beq
{\cal H} = -J \sum_{e_{ij}} \delta_{\sigma_i,\sigma_j} \ , 
\label{hpotts}
\eeq
where $J$ is the spin-spin interaction constant. 
Let us define $\beta = 1/(k_BT)$ (not to be confused with $\beta(G)$), where
$k_B$ is the Boltzmann constant, and denote $K=\beta J$.  Then the partition 
function of this model on a graph $G$ is given by 
\beq 
Z(G,q,v) = \sum_{\{\sigma\}} e^{-\beta {\cal H}} = \sum_{\{\sigma_i\}} \Bigg [ \prod_{e_{ij}}
  e^{K\delta_{\sigma_i,\sigma_j}} \Bigg ] = 
\sum_{\{\sigma\}}\prod_{e_{ij}} \Bigg [ (1+v\delta_{\sigma_i,\sigma_j}) \Bigg ]
\ , 
\label{zpotts}
\eeq
where here $\{ \sigma \}$ denotes the set of all values of the $\sigma$
variables on the vertices of $G$ and 
\beq
v \equiv e^K-1 \ . 
\label{veq}
\eeq
The intervals $v \ge 0$ and $-1 \le v \le 0$ correspond, respectively to the
ferromagnetic and antiferromagnetic Potts models.  The value $v=-1$, i.e.,
$K=-\infty$, corresponds to the zero-temperature Potts antiferromagnet. This
provides an understanding of the relation (\ref{pz}); in the limit $K \to
-\infty$, the only spin configurations that contribute to the Potts model
partition function are those for which $\sigma_i \ne \sigma_j$ on adjacent
vertices $v_i$ and $v_j$ of $G$, and, with the isomorphism between the values
of these $\sigma_i$ and $\sigma_j$ variables and assignments of colors to the
vertices of $G$, these spin configurations are precisely isomorphic to a proper
$q$-coloring of the vertices of $G$.

The dimensionless free energy, per vertex, of the Potts
model on a graph $G$ (usually a regular lattice graph) in the limit $n(G) \to
\infty$, is
\beq
f(\{G\},q,v) = \lim_{n(G) \to \infty} \frac{1}{n(G)} \, \ln[Z(G,q,v)] \ . 
\label{f}
\eeq
The ground-state degeneracy, per vertex, of the $q$-state Potts 
antiferromagnet on a graph $G$ in the limit $n(G) \to \infty$ is 
\beq
W(\{G\},q) = \lim_{n(G) \to \infty} [P(G,q)]^{1/n(G)} \ . 
\label{w}
\eeq
As noted above, in statistical physics, one is commonly interested in integral
values of $q \ge \chi(G)$, is usually a positive integer, but
Eq. (\ref{zcluster}) enables one to generalize $q$ to other values.  Since
$P(G,q)$ and/or $Z(G,q,v)$ may be negative for (real) values of $q$ away from
the positive integers, the evaluation of the relations (\ref{f}) and (\ref{w})
requires specification of which of the $n$ roots of $(-1)$ one uses
\cite{w,a}. However, the magnitudes $|f(\{G\},q,v)|$ and $|W(\{G\},q)|$ are
unambiguously defined by Eqs. (\ref{f}) and (\ref{w}). 
The values $q=-1$ and $q=0$ are relevant for the evaluation of the
exponential growth constants of acyclic orientations of directed edges of $G$
and acyclic orientations of directed edges of $G$ that have a unique source
vertex, as specified in Eqs. (\ref{alpha_wqm1}) and (\ref{alpha0_wq0}).  

The Tutte polynomial of a graph $G$ is defined as
\beq
T(G,x,y) = \sum_{G' \subseteq G} (x-1)^{k(G')-k(G)} \, (y-1)^{c(G')} \ . 
\label{t}
\eeq
Two basic properties that are relevant here are that (i) if $G$ contains a
loop, then it has no acyclic orientations, so $a(G)=T(G,2,0)=0$ and
$a_0(G)=T(G,1,0)=0$; and (ii) if $G$ contains a bridge, then it has no
totally cyclic orientations, so $b(G)=T(G,0,2)=0$.  One can write the Tutte
polynomial of a graph $G$ as
\beq
T(G,x,y) = \sum_{i,j} t_{ij} \, x^i \, y^j \ , 
\label{tij}
\eeq
where the $t_{ij}$ can be determined from the definition (\ref{t}).
A basic property of $T(G,x,y)$ is that the nonzero $t_{ij}$ are positive
(integers) \cite{graphtheory,tutte67}. 

The Tutte polynomial and Potts model partition functions are equivalent, and
are related according to 
\beq
Z(G,q,v) = (x-1)^{k(G)}(y-1)^{n(G)}T(G,x,y) \ , 
\label{zt}
\eeq
with the definitions 
\beq
x = 1 + \frac{q}{v} \ , 
\label{xqv}
\eeq
and
\beq
y=v+1
\label{yv}
\eeq
so that
\beq
q=(x-1)(y-1) \ . 
\label{qxy}
\eeq
Hence, 
\beq
P(G,q) = q^{k(G)}(-1)^{n(G)-k(G)} \, T(G,1-q,0) \ . 
\label{pt}
\eeq
Without loss of generality, we restrict ourselves to connected graphs here, 
i.e., $k(G)=1$, so Eq. (\ref{pt}) reduces to
\beq
P(G,q) = q(-1)^{n(G)-1} \, T(G,1-q,0) \ . 
\label{ptcon}
\eeq
From the representation of the Potts model
partition function $Z(G,q,v)$ as a sum of contributions from spanning
subgraphs, Eq. (\ref{zcluster}), it follows that $Z(G,q,v)$ also has an overall
factor of $q$, so that it one can define a reduced Potts model partition
function that is a polynomial in $q$ and $v$,
\beq
Z_r(G,q,v) \equiv \frac{Z(G,q,v)}{q}
\label{zr}
\eeq
Thus, 
\beq
a_0(G) = (-1)^{n(G)-1} \, P_r(G,0) = (-1)^{n(G)-1}Z_r(G,0,-1) = T(G,1,0)
\label{a0_pq0rel}
\eeq

Let us consider a strip graph of the square or triangular lattices with width
$L_y$ vertices and length $L_x=m$ vertices and with a set of boundary
conditions (BCs) in the longitudinal ($x$) and transverse ($y$) directions. Our
discussion also applies to other lattice strips, such as the honeycomb lattice,
with appropriate modifications. For example, let us consider strip graphs with
periodic longitudinal BCs and free transverse BCs, which we denote as
cyclic. For these cyclic lattice strip graph, $P(G,q)$ has the general form
\cite{saleur,cf}
\beq
P(\Lambda,L_y \times m,cyc,q) = \sum_{d=0}^{L_y} c^{(d)}(q) \, 
\sum_{j=1}^{n_P(L_y,d)} [\lambda_{P,\Lambda,L_y,d,j}(q)]^m \ , 
\label{pcyc}
\eeq
where the $\lambda_{P,\Lambda,L_y,d,j}$ are certain algebraic functions
depending on the type of lattice $\Lambda$ (including transverse 
boundary conditions), 
the strip width $L_y$, and the value of $d$, but not the length $m$. 
Similarly, for cyclic lattice graphs, $Z(G,q,v)$ has the general form 
\beq
Z(\Lambda,L_y \times m,cyc,q,v) = \sum_{d=0}^{L_y} c^{(d)}(q) \, 
\sum_{j=1}^{n_Z(L_y,d)} [\lambda_{Z,\Lambda,L_y,d,j}(q,v)]^m \ , 
\label{zcyc}
\eeq
where again the $\lambda_{Z,\Lambda,L_y,d,j}$ are certain algebraic functions
depending on the type of lattice $\Lambda$ (including transverse boundary
conditions), the strip width $L_y$, and the value of $d$, but not the length
$m$. In the text and tables, to avoid cumbersome notation, we will often omit
the subscripts $P$ and $Z$ and distinguish between the $\lambda$ functions for
$P(G,q)$ and $Z(G,q,v)$ either by context or by their respective arguments $q$
and $(q,v)$.  The coefficients $c^{(d)}(q)$ are polynomials of degree $d$
defined by
\beq
c^{(d)}(q) = \sum_{j=0}^d (-1)^j {2d-j \choose j} \, q^{d-j} \ , 
\label{cd}
\eeq
so $c^{(0)}=1$, $c^{(1)}=q-1$, etc. The numbers $n_P(L_y,d)$ and $n_Z(L_y,d)$
will not be needed here.  Because the factor of $q$ is not manifest in the form
(\ref{pcyc}), the evaluation of $P_r(G,q)$ at $q=0$ that is necessary to obtain
$a_0(G)$ (see Eq.  (\ref{a0_prq0})) requires one to take a limit, $\lim_{q \to
  0} P_r(G,q)$, which one can perform, e.g., by use of L'H\^{o}pital's rule.  A
convenient summary of the quantities that we calculate in the text and their
relation with the chromatic and Tutte polynomials is given in Table
\ref{quantities_table}.

The square, triangular, and honeycomb lattices are special cases of the set of
Archimedean lattices.  An Archimedean lattice is defined as a uniform tiling of
the plane with one or more types of regular polygons, such that all vertices
are equivalent, and hence is $\Delta$-regular.  In general, an Archimedean
lattice $\Lambda$ is identified by the ordered sequence of regular polygons
traversed in a circuit around any vertex: $\Lambda = (\prod p_i^{a_i})$, where
the $i$'th polygon has $p_i$ sides and appears $a_i$ times contiguously in the
sequence (it can also occur non-contiguously).

\end{appendix}



\newpage

\begin{sidewaystable}
  \caption{\footnotesize{Values of $\epsilon$, $\alpha$, $\alpha_0$, $\beta$,
      $\rho_\alpha$, $\rho_{\alpha_0}$, and $\rho_\beta$ for the
      infinite-length limits of some simple strip graphs.  See text for
      notation. The strips are listed in order of increasing vertex degree
      $\Delta$ or $\Delta_{eff}$. Short floating-point evaluations of exact
      expressions are included.}}
\begin{center}
\begin{tabular}{|c|c|c|c|c|c|c|c|c|} \hline\hline
$\{ G \}$ &$\Delta$ or $\Delta_{eff}$& $\epsilon$ & $\alpha$ & $\alpha_0$ 
& $\beta$ & $\rho_\alpha$ & $\rho_{\alpha_0}$ & $\rho_\beta$ \\ 
\hline
$\{ HL\}$ & 5/2 & $2^{5/4}=2.378$&$(31)^{1/4}=2.360$ & $5^{1/4}=1.495$ & 
$\sqrt{2}=1.414$ & $\Big ( \frac{31}{32}\Big )^{1/4}=0.992$ & 
            $\Big ( \frac{5}{32} \Big )^{1/4}=0.629$ & 
            $\Big ( \frac{1}{8}  \Big )^{1/4}=0.595$ \\
$\{ L \}$ &  3  & $2^{3/2}=2.828$&$\sqrt{7}=2.646$&$\sqrt{3}=1.732$& 2 &
$\sqrt{ \frac{7}{8}}=0.935$  & $\sqrt{ \frac{3}{8}}=0.612$  & 
            $\frac{1}{\sqrt{2}}=0.707$ \\
$\{ Wh\}$ & 4 & 4 & 3 & 2 & 3 & $\frac{3}{4}$ & $\frac{1}{2}$ & $\frac{3}{4}$
  \\
$\{TL \}$ & 4 & 4 & 3 & 2 & $\frac{3+\sqrt{13}}{2}=3.303$ & 
 $\frac{3}{4}$ & $\frac{1}{2}$ & $\frac{3+\sqrt{13}}{8}=0.826$ \\
$\{sq_d\}$ & 5 & $2^{5/2}=5.657$ & $2\sqrt{3}=3.464$ & $\sqrt{6}=2.449$ 
& $\sqrt{13+\sqrt{181}}=5.143$ & $\sqrt{\frac{3}{8}}=0.612$ & 
$\frac{\sqrt{3}}{4}=0.433$ & $\sqrt{\frac{13+\sqrt{181}}{32}}=0.909$ \\
\hline\hline
\end{tabular}
\end{center}
\label{strip_table}
\end{sidewaystable}


\begin{table}
  \caption{\footnotesize{Values of $\alpha(\{G\})$ for
      the infinite-length limits of strip graphs of the square lattice with 
      width $L_y$ vertices and free (F) or periodic (P) transverse boundary
      conditions, $BC_y$  Here, as discussed in the text, 
      $\alpha(sq,(L_y)_{BC_y} \times \infty) =
      [\lambda_{sq,(L_y)_{BC_y}}(-1)]^{1/L_y}$, and these values are 
      inferred to be lower bounds on $\alpha(sq)$, with the values for 
      periodic $BC_y$ and the maximal $L_y$ being the most restrictive. Here 
      and in subsequent tables, a blank entry means that the 
      evaluation is not applicable.}}
\begin{center}
\begin{tabular}{||l|l|l|l||}
\hline\hline
BC$_y$ & $L_y$ & $\alpha(sq,(L_y)_{BC_y} \times \infty)$ & 
$R_{sq,BC_y,\frac{L_y}{L_y-1} }$ \\ 
\hline\hline
F & 1  & 2          &  \\ \hline
F & 2  & $\sqrt{7} = 2.64575131$  & 1.32287566 \\ \hline
F & 3  & $\Big ( \frac{27 + \sqrt{481}}{2} \Big)^{1/3} = 2.90304302$ & 
1.09724713 \\ \hline
F & 4  & 3.04073149   & 1.04742901  \\ \hline
F & 5  & 3.12642125   & 1.02818064  \\ \hline
F & 6  & 3.18487566   & 1.01869691  \\ \hline
F & 7  & 3.22729404   & 1.01331934  \\ \hline
F & 8  & 3.25947731   & 1.00997215  \\ 
\hline\hline
P & 3  & $(34)^{1/3} = 3.2396118$  & 1.22445817 \\ \hline
P & 4  & $\Big ( \frac{139 + \sqrt{16009}}{2} \Big )^{1/4} = 3.39445098$  & 
1.04779560 \\ 
\hline
P & 5 & $\Big ( \frac{527 + \sqrt{200585}}{2} \Big )^{1/5} = 3.44812570$ & 
1.01581249 \\
 \hline
P & 6 & 3.47054571    & 1.00650209  \\ \hline
P & 7 & 3.48113984    & 1.00305258  \\ \hline
P & 8 & 3.48658682    & 1.00156471  \\ \hline
P & 9 & 3.48956089    & 1.00085301  \\ \hline
P &10 & 3.49125850    & 1.00048648  \\ \hline
P &11 & 3.49226085    & 1.00028710  \\ \hline
P &12 & 3.49286857    & 1.00017402  \\ \hline
P &13 & 3.493244875   & 1.000107736  \\ \hline 
\end{tabular}
\end{center}
\label{lowerbounds_alpha_sq_table}
\end{table}


\begin{table}
  \caption{\footnotesize{Upper bounds and their ratios for 
$\alpha(sq)$ as functions of strip width $L_y$.}} 
\begin{center}
\begin{tabular}{||l|l|l||}
\hline
$\frac{L_y+1}{L_y}$ & 
$\frac{\lambda_{sq,L_y+1,free}(-1)}{\lambda_{sq,L_y,free}(-1)}$ & 
$R_{sq, \frac{L_y^2}{(L_y-1)(L_y+1)}, free} (-1)$ \\ \hline
2/1  & 3.5         &             \\ \hline
3/2  & $\frac{27 + \sqrt{481}}{14} = 3.49512230$ & 1.00139557 \\ \hline
4/3  & 3.49423306  & 1.00025449  \\ \hline
5/4  & 3.49401836  & 1.00006145  \\ \hline
6/5  & 3.49395589  & 1.00001788  \\ \hline
7/6  & 3.49393533  & 1.00000588  \\ \hline
8/7  & 3.493927961 & 1.000002100  \\ \hline
\end{tabular}
\end{center}
\label{upperbounds_alpha_sq_table}
\end{table}


\begin{table}
  \caption{\footnotesize{Lower bounds and their ratios for 
$\alpha_0(sq)$ as functions of strip width $L_y$. In this table and the others,
the abbreviation cyl stands for ``cylindrical''.}} 
\begin{center}
\begin{tabular}{||l|l|l|l||}
\hline
BC          & $L_y$ & $[\lambda_{sq,L_y,free/cyl}(0)]^{1/L_y}$ & 
$R_{sq, \frac{L_y}{L_y-1}, free/cyl} (0)$ \\ \hline
free        & 1   & 1        & \\ \hline
free        & 2   & $\sqrt{3} = 1.73205081$            & 1.73205081 \\ \hline
free        & 3   & $(5+\sqrt{14}\, )^{1/3}=2.05998754$ & 1.18933436 \\ \hline
free        & 4   & 2.24131157 & 1.08802190 \\ \hline
free        & 5   & 2.35572295 & 1.05104662 \\ \hline
free        & 6   & 2.43432494 & 1.03336640 \\ \hline
free        & 7   & 2.49159809 & 1.02352733 \\ \hline
free        & 8   & 2.53516365 & 1.01748499 \\ \hline \hline
cyl & 3  & $(13)^{1/3} = 2.35133469$ & 1.35754371 \\ \hline
cyl & 4  & $(23+2\sqrt{111}\, )^{1/4} = 2.57655243$ & 1.09578294 \\ \hline
cyl & 5  & $(74+11\sqrt{34}\, )^{1/5} = 2.67956432$ & 1.03998052 \\ \hline
cyl & 6          & 2.73462860 & 1.02054971 \\ \hline
cyl & 7          & 2.76735961 & 1.01196909 \\ \hline
cyl & 8          & 2.78834612 & 1.00758359 \\ \hline
cyl & 9          & 2.80258484 & 1.00510651 \\ \hline
cyl & 10         & 2.81267772 & 1.00360127 \\ \hline
cyl & 11         & 2.82008605 & 1.00263390 \\ \hline
cyl & 12         & 2.82568101 & 1.00198397 \\ \hline
cyl & 13         & 2.830007783& 1.001531233 \\ \hline
\end{tabular}
\end{center}
\label{lowerbounds_alpha0_sq_table}
\end{table}


\begin{table}
  \caption{\footnotesize{Upper bounds and their ratios for 
$\alpha_0(sq)$ as functions of strip width $L_y$.}} 
\begin{center}
\begin{tabular}{||l|l|l||}
\hline
$\frac{L_y+1}{L_y}$ & 
$\frac{\lambda_{sq, L_y+1, free} (0)}{\lambda_{sq, L_y, free} (0)}$ & 
$R_{sq, \frac{L_y^2}{(L_y-1)(L_y+1)}, free} (0)$ \\ \hline
2/1 & 3                 &  \\ \hline
3/2& $\frac{5+\sqrt{14}}{3}=2.91388580$ & 1.02955305 \\ 
\hline
4/3 & 2.88678970 & 1.00938624 \\ \hline
5/4 & 2.87482980 & 1.00416021 \\ \hline
6/5 & 2.86846939 & 1.00221735 \\ \hline
7/6 & 2.86467029 & 1.00132619 \\ \hline
8/7 & 2.862213752& 1.000858265 \\ \hline
\end{tabular}
\end{center}
\label{upperbounds_alpha0_sq_table}
\end{table}


\begin{table}
  \caption{\footnotesize{Lower bounds on $\alpha (tri)$ and their ratios
      relative to the exact value (\ref{alpha_tri}), as functions of strip
width $L_y$.}}
\begin{center}
\begin{tabular}{||l|l|l|l|l||}
\hline
BC     & $L_y$ & $[\lambda_{tri,L_y,free/cyl}(-1)]^{1/L_y}$ &
$\frac{[\lambda_{tri,L_y,free/cyl}(-1)]^{1/L_y}}{\alpha(tri)}$ &
$R_{tri, \frac{L_y}{L_y-1}, free/cyl} (-1)$ \\ \hline
free       & 2     & 3                 & 0.67044390 & \\ \hline
free       & 3     & $\Big ( \frac{43 + \sqrt{1417}}{2} \Big )^{1/3}$ &
0.766337701111...\ & 1.143030310705... \\
 & & $= 3.429090932116...$ & & \\ \hline
free       & 4     & 3.66535037 & 0.81913727 & 1.06889856 \\ \hline
free       & 5     & 3.81466660 & 0.85250665 & 1.04073723 \\ \hline
free       & 6     & 3.91752078 & 0.87549264 & 1.02696282 \\ \hline
free       & 7     & 3.99266294 & 0.89228551 & 1.01918105 \\ \hline
free       & 8     & 4.04995674 & 0.90508960 & 1.01434977 \\ \hline
free       & 9     & 4.09508340 & 0.91517457 & 1.01114251 \\ \hline \hline
cyl & 2   & $2\sqrt{3} = 3.46410162$ & 0.77416193 & \\ \hline
cyl & 3   & $(71)^{1/3} = 4.14081775$ & 0.92539534 & 1.19535112 \\ \hline
cyl & 4   & $(2^5 \times 11)^{1/4}$ & 0.96800334 & 1.04604303 \\
 & & $ = 4.33147354$ & & \\ \hline
cyl & 5   & $[3(299+113\sqrt{5} \, )]^{1/5}$ & 0.98401611 & 1.01654206 \\
 & & $ = 4.40312504$ & & \\ \hline
cyl & 6     & 4.43528747 & 0.99120381 & 1.00730445  \\ \hline
cyl & 7     & 4.45150713 & 0.99482860 & 1.00365696  \\ \hline
cyl & 8     & 4.46037926 & 0.99681136 & 1.00199306  \\ \hline
cyl & 9     & 4.46552972 & 0.99796239 & 1.00115471  \\ \hline
cyl & 10    & 4.46865768 & 0.99866143 & 1.00070047 \\ \hline
cyl & 11    & 4.47062537 & 0.99910117 & 1.00044033 \\ \hline
cyl & 12    & 4.471898356 & 0.999385660 & 1.0002847452 \\ \hline
\end{tabular}
\end{center}
\label{lowerbounds_alpha_tri_table}
\end{table}


\begin{table}
  \caption{\footnotesize{Upper bounds on $\alpha(tri)$, their ratios
      relative to the exact value (\ref{alpha_tri}), and ratios of adjacent
      upper bounds, as functions of strip 
width $L_y$.}} 
\begin{center}
\begin{tabular}{||l|l|l|l||}
\hline
$\frac{L_y+1}{L_y}$ & $\frac{\lambda_{tri, L_y+1, free} (-1)}
{\lambda_{tri, L_y, free} (-1)}$ & 
$\frac{\lambda_{tri, L_y+1, free} (-1)/\lambda_{tri, L_y, free} (-1)}
{\alpha(tri)}$ & $R_{tri, \frac{L_y^2}{(L_y-1)(L_y+1)}, free} (-1)$ \\ \hline
2/1 & 4.5               & 1.00566585 & \\ \hline
3/2 & $\frac{43 + \sqrt{1417}}{18} = 4.48017002$ & 1.00123422 & 1.00442617 \\ \hline
4/3 & 4.47635966 & 1.00038268   & 1.00085122 \\ \hline
5/4 & 4.47528766 & 1.00014311   & 1.00023954 \\ \hline
6/5 & 4.47491635 & 1.000060125  & 1.00008298 \\ \hline
7/6 & 4.47476968 & 1.00002735   & 1.00003278 \\ \hline
8/7 & 4.47470626 & 1.000013175  & 1.00001417 \\ \hline
9/8 & 4.474676977& 1.000006630  & 1.0000065451 \\ \hline
\end{tabular}
\end{center}
\label{upperbounds_alpha_tri_table}
\end{table}


\begin{table}
  \caption{\footnotesize{Lower bounds on $\alpha_0(tri)$ and their ratios
      relative to the exact value (\ref{alpha0_tri}), and ratios of adjacent
      bounds, as 
      functions of strip width $L_y$.}} 
\begin{center}
\begin{tabular}{||l|l|l|l|l||}
\hline
BC     & $L_y$ & $[\lambda_{tri, L_y, free/cyl}(0)]^{1/L_y}$ & 
$\frac{[\lambda_{tri, L_y, free/cyl}(0)]^{1/L_y}}{\alpha_0(tri)}$ & 
$R_{tri, \frac{L_y}{L_y-1}, free/cyl} (0)$ \\ \hline
free      & 2    & 2           & 0.53037459 &  \\ \hline
free      & 3    & $\Big ( \frac{17 + \sqrt{193}}{2} \Big )^{1/3}$ 
& 0.66043002 & 1.24521429 \\ 
 & & $ = 2.49042857$ & & \\ \hline
free      & 4    & 2.77154840 & 0.73497943 & 1.11288010 \\ \hline
free      & 5    & 2.95242249 & 0.78294494 & 1.06526102 \\ \hline
free      & 6    & 3.07822224 & 0.81630543 & 1.04260899 \\ \hline
free      & 7    & 3.17066700 & 0.84082061 & 1.03003187 \\ \hline
free      & 8    & 3.24142502 & 0.85958474 & 1.02231645 \\ \hline
free      & 9    & 3.29730594 & 0.87440365 & 1.01723962 \\ \hline \hline
cyl & 2  & $\sqrt{6} = 2.44948974$ & 0.64957356 & \\ \hline
cyl & 3  & $2^{5/3} =  3.17480210$ & 0.84191719 & 1.29610753 \\ \hline
cyl & 4  & $[6(12+\sqrt{129} \, )]^{1/4}$ & 0.91242797& 1.08375026 \\ 
 & & $ = 3.440692605$ & & \\ \hline
cyl & 5     & $(307 + 29\sqrt{85} \, )^{1/5}$ & 0.94491029 & 1.03559988 \\ 
 & & $ = 3.56318084$ & & \\ \hline
cyl & 6     & 3.628852235 & 0.96232551 & 1.01843055 \\ \hline
cyl & 7     & 3.667909685 & 0.97268305 & 1.01076303 \\ \hline
cyl & 8     & 3.69293928  & 0.97932058 & 1.00682394 \\ \hline
cyl & 9     & 3.70990510  & 0.98381970 & 1.00459412 \\ \hline
cyl & 10    & 3.72191820  & 0.98700543 & 1.003238115 \\ \hline
cyl & 11    & 3.73072654  & 0.98934128 & 1.00236661  \\ \hline
cyl & 12    & 3.737371971 & 0.991103569& 1.001781271 \\ \hline
\end{tabular}
\end{center}
\label{lowerbounds_alpha0_tri_table}
\end{table}


\begin{table}
  \caption{\footnotesize{Upper bounds on $\alpha_0(tri)$, their ratios
      relative to the exact value (\ref{alpha0_tri}), and ratios of adjacent
      bounds, as functions of strip 
width $L_y$.}} 
\begin{center}
\begin{tabular}{||l|l|l|l||}
\hline
$\frac{L_y+1}{L_y}$ & 
$\frac{\lambda_{tri,L_y+1,free}(0)}{\lambda_{tri,L_y,free}(0)}$ & 
$\frac{\lambda_{tri,L_y+1,free}(0)/ \lambda_{tri,L_y,free}(0)}{\alpha_0(tri)}$
& $R_{tri, \frac{L_y^2}{(L_y-1)(L_y+1)},free} (0)$ \\ \hline
2/1 & 4                 & 1.06074919 & \\ \hline
3/2 & $\frac{17 + \sqrt{193}}{8} = 3.86155550$ & 1.02403546 & 1.03585200
\\ \hline
4/3 & 3.82003723  & 1.01302535 & 1.01086855 \\ \hline
5/4 & 3.80191720  & 1.00822014 & 1.00476603 \\ \hline
6/5 & 3.79234033  & 1.00568048 & 1.00252532 \\ \hline
7/6 & 3.78664508  & 1.00417017 & 1.00150404 \\ \hline
8/7 & 3.78297452  & 1.003196785& 1.00097028 \\ \hline
9/8 & 3.780466270 & 1.002531630& 1.000663476 \\ \hline
\end{tabular}
\end{center}
\label{upperbounds_alpha0_tri_table}
\end{table}


\begin{table}
  \caption{\footnotesize{Lower bounds on $\alpha(hc)$ and their ratios,
      as functions of strip width $L_y$.}} 
\begin{center}
\begin{tabular}{||l|l|l|l||}
\hline
BC          & $L_y$ & $[\lambda_{hc, L_y, free/cyl}(-1)]^{1/(2L_y)}$ & 
$R_{hc, \frac{L_y}{L_y-1} / \frac{L_y}{L_y-2}, free/cyl} (-1)$ \\ \hline
free    & 2     & $(31)^{1/4} = 2.35961106$ & \\ \hline
free    & 3     & 2.49321528 & 1.05662129 \\ \hline
free    & 4     & 2.56281578 & 1.02791596 \\ \hline
free    & 5     & 2.60550411 & 1.01665681 \\ \hline
free    & 6     & 2.63435715 & 1.01107388 \\ \hline \hline
cyl     & 2     & $\sqrt{7} = 2.64575131$ & \\ \hline
cyl     & 4     & 2.77349764 & 1.04828357 \\ \hline
cyl     & 6     & 2.782197008& 1.003136606 \\ \hline
\end{tabular}
\end{center}
\label{lowerbounds_alpha_hc_table}
\end{table}


\begin{table}
  \caption{\footnotesize{Upper bounds on $\alpha(hc)$ and their ratios,
      as functions of strip width $L_y$.}} 
\begin{center}
\begin{tabular}{||l|l|l||}
\hline
$(L_y+1)/L_y$ & 
$\sqrt{\lambda_{hc,L_y+1,free}(-1)/\lambda_{hc,L_y,free}(-1)}$ & 
$R_{hc, \frac{L_y^2}{(L_y-1)(L_y+1)}, free} (-1)$ \\ \hline
2/1     & $\frac{\sqrt{31}}{2} = 2.78388218$ & \\ \hline
3/2     & 2.78354659 & 1.00012056 \\ \hline
4/3     & 2.78349352 & 1.00001907 \\ \hline
5/4     & 2.78348737 & 1.00000221 \\ \hline
6/5     & 2.783486470& 1.000000323 \\ \hline
\end{tabular}
\end{center}
\label{upperbounds_alpha_hc_table}
\end{table}


\begin{table}
  \caption{\footnotesize{Lower bounds on $\alpha_0(hc)$ and their ratios,
      as functions of strip width $L_y$.}} 
\begin{center}
\begin{tabular}{||l|l|l|l||}
\hline
BC     & $L_y$ & $[\lambda_{hc, L_y, free/cyl}(0)]^{1/(2L_y)}$ & 
$R_{hc,\frac{L_y}{L_y-1} / \frac{L_y}{L_y-2}, free/cyl} (0)$ \\ \hline
free   & 2     & $5^{1/4} = 1.49534878$ & \\ \hline
free   & 3     & 1.69793365 & 1.13547667 \\ \hline
free   & 4     & 1.80571700 & 1.06347913 \\ \hline
free   & 5     & 1.87241553 & 1.03693742 \\ \hline
free   & 6     & 1.91770572 & 1.02418811 \\ \hline \hline
cyl   & 2     & $\sqrt{3} = 1.73205081$ &  \\ \hline
cyl   & 4     & 2.04591494 & 1.18120954 \\ \hline
cyl   & 6     & 2.106218408 &1.029475062 \\ \hline
\end{tabular}
\end{center}
\label{lowerbounds_alpha0_hc_table}
\end{table}


\begin{table}
  \caption{\footnotesize{Upper bounds on $\alpha_0(hc)$ and their ratios,
      as functions of strip width $L_y$.}} 
\begin{center}
\begin{tabular}{||l|l|l||}
\hline
$(L_y+1)/L_y$ & $\sqrt{\lambda_{hc, L_y+1, free} (0) / \lambda_{hc, L_y, free} (0)}$ & $R_{hc, \frac{L_y^2}{(L_y-1)(L_y+1)}, free} (0)$ \\ \hline
2/1 & $\sqrt{5} = 2.23606798$ & \\ \hline
3/2 & 2.18915819 & 1.02142823 \\ \hline
4/3 & 2.17188387 & 1.00795361 \\ \hline
5/4 & 2.16477332 & 1.00328466 \\ \hline
6/5 & 2.161128567 &1.001686502 \\ \hline
\end{tabular}
\end{center}
\label{upperbounds_alpha0_hc_table}
\end{table}


\begin{table}
  \caption{\footnotesize{Lower bounds on $\alpha(sq_d)$ and their ratios,
      as functions of strip width $L_y$.}} 
\begin{center}
\begin{tabular}{||l|l|l|l||}
\hline
BC      & $L_y$ & $[\lambda_{sq_d, L_y, free/cyl}(-1)]^{1/L_y}$ & 
$R_{sq_d, \frac{L_y}{L_y-1}, free/cyl} (-1)$ \\ \hline
free    & 2   & $2\sqrt{3} = 3.464101615$ & \\ \hline
free    & 3   & $[4(9 + \sqrt{69} \, )]^{1/3}=4.10604888$ & 1.18531421 \\ 
\hline
free    & 4     & 4.46677215 & 1.08785167 \\ \hline \hline
cyl & 3     & $2 \times (15)^{1/3} = 4.93242415$ & \\ \hline
cyl & 4     & 5.354782509 & 1.085628962 \\ \hline
\end{tabular}
\end{center}
\label{lowerbounds_alpha_sqd_table}
\end{table}


\begin{table}
  \caption{\footnotesize{Lower bounds on $\alpha_0(sq_d)$ and their ratios,
      as functions of strip width $L_y$.}} 
\begin{center}
\begin{tabular}{||l|l|l|l||}
\hline
BC   & $L_y$ & $[\lambda_{sq_d, L_y, free/cyl}(0)]^{1/L_y}$ & 
$R_{sq_d, \frac{L_y}{L_y-1}, free/cyl} (0)$ \\ \hline
free   & 2  & $\sqrt{6} = 2.44948974$ & \\ \hline
free   & 3 & $\big [ \frac{3(11+\sqrt{97}\, )}{2} \big ]^{1/3} = 3.15058481$ & 
 1.28622086 \\ \hline
free   & 4  & 3.55858048 & 1.12949839 \\ \hline \hline
cyl & 3    & $(60)^{1/3} = 3.91486764$ & \\ \hline
cyl & 4    & 4.417285760 & 1.128335915 \\ \hline
\end{tabular}
\end{center}
\label{lowerbounds_alpha0_sqd_table}
\end{table}


\clearpage

\begin{table}
  \caption{\footnotesize{Lower bounds and their ratios for $\beta(sq)$ as 
functions of strip width $L_y$. The abbreviations cycl and tor stand for
``cyclic'' and ``toroidal'', respectively.}} 
\begin{center}
\begin{tabular}{||l|l|l|l||}
\hline
BC  & $L_y$ & $[\lambda_{sq, L_y, cyc/tor}(-1,1)]^{1/L_y}$ & 
$R_{sq, \frac{L_y}{L_y-1}, cyc/tor} (-1,1)$ \\ \hline
cycl  & 1   & 1 &  \\ \hline
cycl  & 2   & 2 & 2 \\ \hline
cycl  & 3   & $\Big( \frac{17 + \sqrt{145}}{2} \Big)^{1/3} = 2.43966477$  & 
1.21983238 \\ \hline
cycl  & 4   & 2.68228611 & 1.09944864 \\ \hline
cycl  & 5   & 2.83465313 & 1.05680491 \\ \hline \hline
tor   & 2     & $\sqrt{10} = 3.16227766$ &  \\ \hline
tor   & 3     & $\Big( \frac{41+\sqrt{1345}}{2} \Big)^{1/3} = 3.38648385$ & 
1.07090022 \\ \hline
tor & 4     & 3.449673447 & 1.018659353 \\ \hline
\end{tabular}
\end{center}
\label{lowerbounds_beta_sq_table}
\end{table}


\begin{table}
  \caption{\footnotesize{Upper bounds and their ratios for $\beta(sq)$ as 
functions of strip width $L_y$.}} 
\begin{center}
\begin{tabular}{||l|l|l||}
\hline
$(L_y+1)/L_y$ & 
$\lambda_{sq,L_y+1,cyc}(-1,1)/\lambda_{sq,L_y,cyc}(-1,1)$ & 
$R_{sq, \frac{L_y^2}{(L_y-1)(L_y+1)}, cyc} (-1,1)$ \\ \hline
2/1 & 4            &    \\ \hline
3/2 & $\frac{17 + \sqrt{145}}{8} = 3.63019932$ & 1.10186787 \\ \hline
4/3 & 3.56475709 & 1.01835812 \\ \hline
5/4 & 3.535730951 &1.0082093749 \\ \hline
\end{tabular}
\end{center}
\label{upperbounds_beta_sq_table}
\end{table}


\begin{table}
  \caption{\footnotesize{Lower bounds and their ratios for $\beta(tri)$ as 
functions of strip width $L_y$.}} 
\begin{center}
\begin{tabular}{||l|l|l|l||}
\hline
BC     & $L_y$ & $[\lambda_{tri, L_y, cyc/tor}(-1,1)]^{1/L_y}$ & 
$R_{tri, \frac{L_y}{L_y-1}, cyc/tor} (-1,1)$ \\ \hline
cycl.  & 2     & $\frac{3+\sqrt{13}}{2} = 3.302775637731...$ & \\ \hline
cycl.  & 3     & 4.48070229 & 1.35664749 \\ \hline
cycl.  & 4     & 5.16971535 & 1.15377345 \\ \hline
cycl.  & 5     & 5.61764092 & 1.08664414 \\ \hline \hline
tor.   & 2     & $\sqrt{2(14+\sqrt{202})} = 7.51168029$ & \\ \hline
tor.   & 3     & 7.696127303 & 1.024554694 \\ \hline
\end{tabular}
\end{center}
\label{lowerbounds_beta_tri_table}
\end{table}


\begin{table}
  \caption{\footnotesize{Upper bounds and their ratios for $\beta(tri)$ as 
functions of strip width $L_y$.}} 
\begin{center}
\begin{tabular}{||l|l|l||}
\hline
$(L_y+1)/L_y$ & 
$\lambda_{tri, L_y+1, cyc} (-1,1) / \lambda_{tri, L_y, cyc} (-1,1)$ & 
$R_{tri, \frac{L_y^2}{(L_y-1)(L_y+1)}, cyc} (-1,1)$ \\ \hline
2/1 & $\frac{11+3\sqrt{13}}{2} = 10.90832691$ &   \\ \hline
3/2 & 8.24669860 & 1.32275077 \\ \hline
4/3 & 7.94014180 & 1.03860848 \\ \hline
5/4 & 7.832553170 &1.013736086 \\ \hline
\end{tabular}
\end{center}
\label{upperbounds_beta_tri_table}
\end{table}


\begin{table}
  \caption{\footnotesize{Lower bounds, their ratios with respect to the exact
      value of $\beta(hc)$, and ratios of adjacent bounds, as 
functions of strip width $L_y$.}} 
\begin{center}
\begin{tabular}{||l|l|l|l|l||}
\hline
BC  & $L_y$ & $[\lambda_{hc, L_y, cyc/tor}(-1,1)]^{1/(2L_y)}$ & 
$\frac{[\lambda_{hc, L_y, cyc/tor}(-1,1)]^{1/(2L_y)}}{\beta(hc)}$ & 
$R_{hc, \frac{L_y}{L_y-1} / \frac{L_y}{L_y-2}, cyc/tor} (-1,1)$ \\ \hline
cycl.  & 2  & $\sqrt{2} = 1.41421356$ & 0.66855262 &  \\ \hline
cycl.  & 3     & 1.62353902 & 0.76750873 & 1.14801545 \\ \hline
cycl.  & 4     & 1.73776398 & 0.82150722 & 1.07035554 \\ \hline
cycl.  & 5     & 1.80926267 & 0.85530738 & 1.04114408 \\ \hline \hline
tor.   & 2     & 2          & 0.94547618 &          \\ \hline
tor.   & 4     & 2.094446760 & 0.990124758 & 1.047223380 \\ \hline
\end{tabular}
\end{center}
\label{lowerbounds_beta_hc_table}
\end{table}


\begin{table}
  \caption{\footnotesize{Upper bounds, their ratios relative to the exact
      $\beta(hc)$, and ratios of adjacent bounds, as 
functions of strip width $L_y$.}} 
\begin{center}
\begin{tabular}{||l|l|l|l||}
\hline
$(L_y+1)/L_y$ & 
$\sqrt{\frac{\lambda_{hc,L_y+1,cyc}(-1,1)}{\lambda_{hc,L_y,cyc}(-1,1)}}$ & 
$\frac{\sqrt{\lambda_{hc,L_y+1,cyc}(-1,1)/\lambda_{hc,L_y,cyc}(-1,1)}} 
{\beta (hc)}$ & $R_{hc, \frac{L_y^2}{(L_y-1)(L_y+1)}, cyc} (-1,1)$ \\ \hline
3/2 & 2.13972616 & 1.01153005 &   \\ \hline
4/3 & 2.13095839 & 1.0073820  &  1.00411447 \\ \hline
5/4 & 2.125910381 & 1.004998810 &1.002374517 \\ \hline
\end{tabular}
\end{center}
\label{upperbounds_beta_hc_table}
\end{table}


\begin{table}
\caption{\footnotesize{Values of $EGC_{ap}(\Lambda)$ defined in Eq.
 (\ref{egc_value})
 with Eqs. (\ref{egc_average}) and (\ref{egc_delta}) for honeycomb, square,
 and triangular lattices, where EGC denotes exponential growth constant. 
 In the cases where we have obtained exact values, namely $\alpha(tri)$,
 $\alpha_0(tri)$, and $\beta(hc)$, these are listed instead of the 
$EGC_{ap}(\Lambda)$ quantity. For reference, we also list the (exactly known)
values of $\tau(\Lambda)$ for these lattices. 
See text for further discussion.}}
\begin{center}
\begin{tabular}{|c|c|c|c|c|c|} \hline\hline
$\Lambda$ & $\Delta(\Lambda)$ & $\alpha(\Lambda)$ & $\alpha_0(\Lambda)$ & 
$\beta(\Lambda)$ & $\tau(\Lambda)$ \\
\hline
hc  & 3 & $2.78284 \pm 0.00064$ & $2.134 \pm 0.027$  
    & 2.11533621655 & 2.24266494889 \\ \hline
sq  & 4 & $3.49359 \pm 0.00034$ & $2.846 \pm 0.016$    
    & $3.49359 \pm 0.00034$ & 3.20991230073 \\ \hline
tri & 6 & 4.47464730907 & 3.77091969752 & $7.7442 \pm 0.0036$ & 
       5.02954607297  \\
\hline\hline
\end{tabular}
\end{center}
\label{egc_values_table}
\end{table}


\begin{table}
\caption{\footnotesize{Values of $\epsilon(\Lambda)$ and 
$\rho_{EGC}(\Lambda)$, defined in (\ref{epsilon}) and
 (\ref{rho_alpha})-(\ref{rho_beta}), 
for honeycomb, square, and triangular lattices $\Lambda$, with the 
 exponential growth constants (EGCs) $\alpha(\Lambda)$, 
$\alpha_0(\Lambda)$, $\beta(\Lambda)$, and $\tau(\Lambda)$. For the EGCs, we
use the exact values of $\rho_\alpha(tri)$, $\rho_{\alpha_0}(tri)$, and 
$\rho_\beta(hc)$ that we have
presented here; for the other EGCs, we use $EGC_{ave}(\Lambda)$ from 
Eq. (\ref{egc_average}) as in Eq. (\ref{rhoest}). 
See text for further discussion.}}
\begin{center}
\begin{tabular}{|c|c|c|c|c|c|} \hline\hline
$\Lambda$ & $\Delta(\Lambda)$ & $\epsilon(\Lambda)$ &
$\rho_\alpha(\Lambda)$ & $\rho_{\alpha_0}(\Lambda)$ & $\rho_\beta(\Lambda)$ \\
\hline
hc  & 3 & $2\sqrt{2}$ & 0.984         & 0.7545       & 0.7478842912 \\ \hline
sq  & 4 & 4           & 0.873         & 0.7115       & 0.873        \\ \hline
tri & 6 & 8           & 0.5593309136  & 0.4713649622 & 0.968         \\
\hline\hline
\end{tabular}
\end{center}
\label{rho_values_table}
\end{table}


\begin{table}
\caption{\footnotesize{Graph-theoretic numbers $a(G)$, $a_0(G)$, and $b(G)$ 
and their expressions as valuations of the chromatic polynomial $P(G,q)$, the
reduced chromatic polynomial $P_r(G,q)=q^{-1} \, P(G,q)$, the Tutte polynomial 
$T(G,x,y)$, and/or the Potts polynomial $Z(G,q,v)$.}}
\begin{center}
\begin{tabular}{|c|c|c|c|c|c|c|} \hline\hline
quantity& $x$ & $y$ & $q$ & $v$  & expression \\ \hline
$a(G)$  &  2  & 0   &$-1$ & $-1$ & $a(G)=T(G,2,0)=(-1)^n \, P(G,-1)$ \\ \hline
$a_0(G)$&  1  & 0   &  0  & $-1$ & $a_0(G)=T(G,1,0)=(-1)^{n-1} \, P_r(G,0)$ \\ \hline
$b(G)$  &  0  & 2   & $-1$&  1   & $b(G)=T(G,0,2)=-Z(G,-1,1)$  \\
\hline\hline
\end{tabular}
\end{center}
\label{quantities_table}
\end{table}


\end{document}